\documentclass[aps,prb,twocolumn,longbibliography,amsfonts,citeautoscript,superscriptaddress,nofootinbib,floatfix,maxnames=4,10pt]{revtex4-2} 
\usepackage[utf8]{inputenc}
\usepackage[english]{babel}
\usepackage{braket}
\usepackage{graphicx}
\usepackage{MnSymbol}
\usepackage{xcolor}
\usepackage{epstopdf}
\usepackage{amsmath} 
\usepackage[normalem]{ulem} %strikethrough \sout 
\usepackage{enumerate}
\usepackage{cases}
\usepackage{lipsum}  

\usepackage[colorlinks=true,linktoc=page,linkcolor=red,citecolor=red,urlcolor=blue]{hyperref}
\usepackage{orcidlink}
\allowdisplaybreaks
\newcommand{\vect}[1]{\boldsymbol{\mathrm{#1}}}
\mathchardef\mhyphen="2D

\newcommand{\NPH}[1]{$N_{ph}=#1$}
\def\-{\sout}
\newcommand{\non}{\nonumber}  

\begin{document}
\title{Topology and energy dependence of Majorana bound states in a photonic cavity}

\author{Aksel Kobiałka~\orcidlink{0000-0003-2881-8253}}
\email[e-mail:]{aksel.kobialka@physics.uu.se}
\affiliation{Department of Physics and Astronomy, Uppsala University, Box 516, S-751 20 Uppsala, Sweden}

\author{Arnob Kumar Ghosh~\orcidlink{0000-0003-0990-8341}}
\email[e-mail:]{arnob.ghosh@physics.uu.se}
\affiliation{Department of Physics and Astronomy, Uppsala University, Box 516, S-751 20 Uppsala, Sweden}

\author{Rodrigo Arouca~\orcidlink{0000-0003-4214-1437}}
\email[e-mail:]{rodrigo-arouca@cbpf.br}
\affiliation{Department of Physics and Astronomy, Uppsala University, Box 516, S-751 20 Uppsala, Sweden}
\affiliation{Centro Brasileiro de Pesquisas Físicas, Rua Doutor Xavier Sigaud 150, Rio de Janeiro,  22290-180, Brazil}

\author{Annica M. Black-Schaffer~\orcidlink{0000-0002-4726-5247}}
%\email[e-mail:]{annica.black-schaffer@physics.uu.se}
\affiliation{Department of Physics and Astronomy, Uppsala University, Box 516, S-751 20 Uppsala, Sweden}
\date{\today}   

%%%%%%%%%%%%%%%%%%%%%%%%%%%%%%%%%%%%%%%%%%%%%%%%%%
\begin{abstract}
Light-matter interaction plays a crucial role in modifying the properties of quantum materials. 
In this work, we investigate the effect of cavity induced photon fields on a topological superconductor hosting Majorana bound states~(MBS). 
We model the system using a Peierls substitution of the photonic operator in the kinetic and spin-orbit terms, and utilize an exact diagonalization of Hamiltonian for a finite number of photons to investigate the coupled system.
We find that the MBS persist even in the presence of a cavity field and notably appear at finite and tunable energy, in contrast to a usual 1D topological superconductor. The MBS energy is shifted by two processes: the cavity photon energy adds a constant energy shift, while the light-matter interaction induces additional parameter dependencies, such that the MBS experience a \textit{pseudo-dispersion} as a function of both light-matter interaction and magnetic field.
Additionally, we find that the MBS energy oscillations are suppressed with increasing light-matter interaction and that disorder stability is not impacted by the light-matter interaction.
Combined, these offer additional tunability and stability of the MBS.
As a second result, we establish a modified spectral localizer formalism as an essential tool for topological characterization of quantum matter in a cavity. 
The spectral localizer allows characterization at arbitrary energies, which is needed for probing different photon sectors. 
However, hybridization between different photon sectors in the low-frequency regime limits a straightforward application of a standard spectral localizer. 
We fully resolve this issue by judiciously applying an energy shift to the spectral localizer.
Our work thus introduces a new avenue for controlling MBS via light–matter coupling and provides a framework for exploring cavity-modified topologies.
\end{abstract}

\maketitle

%===============================
\section{Introduction}
%===============================

Topological superconductors hosting Majorana bound states~(MBS) have emerged as one of the most vibrant areas of research in condensed matter physics~\cite{Kitaev_2001,qi2011topological,Alicea_2012,beenakker2013search,ramonaquado2017,tanaka2024theory}. 
MBS are fascinating states obeying non-abelian statistics, generating fractional charge and thermal conductance~\cite{akhmerov09,akhmerov11,grosfeld11,vaezi13,sticlet14}, as well as being proposed as crucial element of fault-tolerant topological quantum computation~\cite{Ivanov2001,Freedman2003topological,NayakRMP2008}. 
The initial proposal of MBS was put forward by Kitaev in a one dimensional $p$-wave superconductor~\cite{Kitaev_2001}, which, however, is scarce in real materials. 
Nevertheless, this Kitaev model of a one dimensional topological superconductor~(1DTSC) has been shown to be realized in heterostructure platforms, most prominently a semiconducting nanowire with strong Rashba spin-orbit coupling~(SOC) in proximity to a conventional bulk $s$–wave superconductor and in an applied external magnetic field~\cite{Oreg2010,LutchynPRL2010,Alicea_2012,Mourik2012Science,das2012zero,ramonaquado2017}. 
While the `smoking-gun' experimental signature of the MBS in such a heterostructure setup is still ambiguous~\cite{Mourik2012Science,das2012zero,DengNano2012,Rokhinson2012,Finck2013,Albrecht2016,Deng2016,NichelePRL2017,JunSciAdv2017,Zhang2017NatCommun,Gul2018,Grivnin2019,ChenPRL2019}, recent achievements in theory and fabrication using what is known as poor man's MBS have shown great promise~\cite{LeijnsePM2012,Dvir23,Haaf24,Zatelli24}.

\begin{figure}[httb!]
    \centering
    \includegraphics[width=0.9\linewidth]{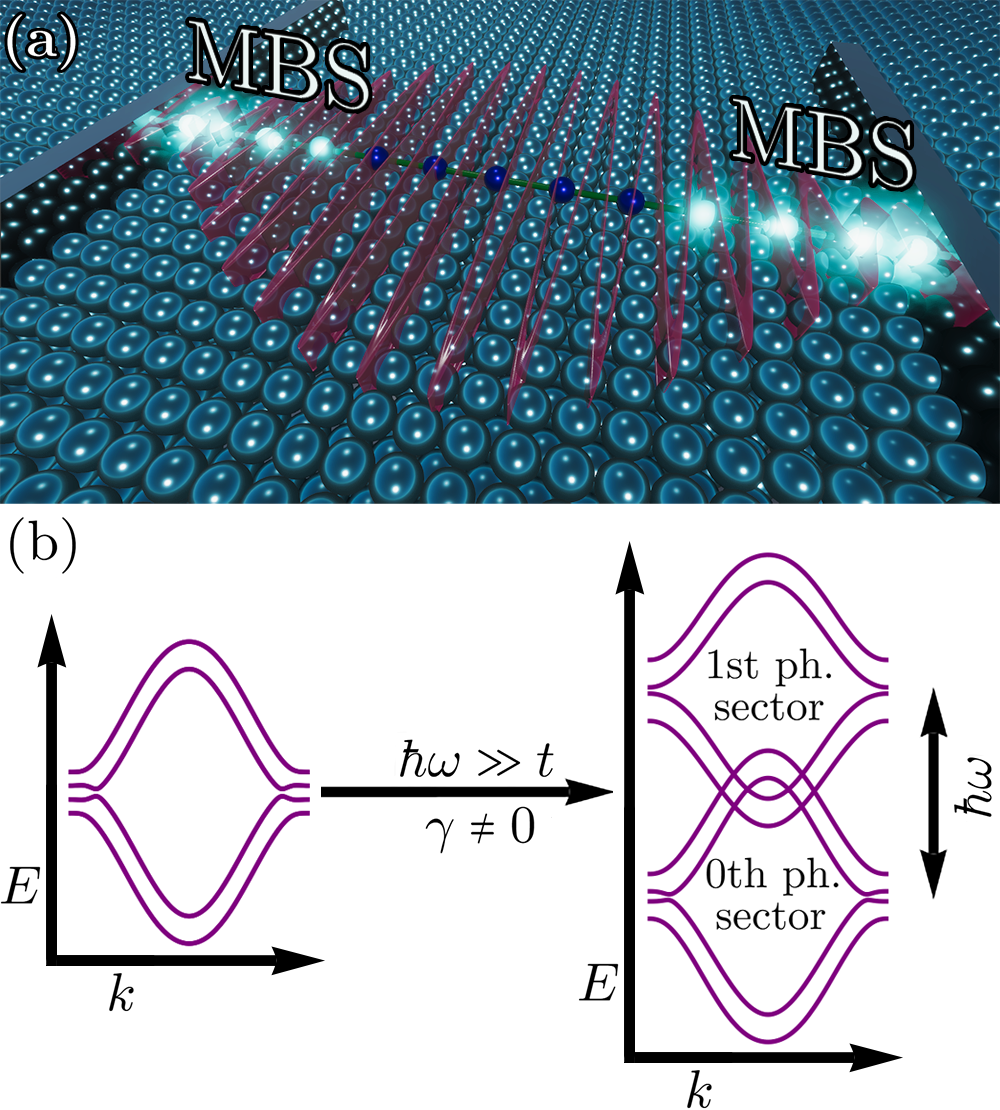}
    \caption{(a) Schematic view of a 1DTSC with light (dark) blue balls denoting the superconducting substrate (Rashba nanowire) exhibiting MBS at its ends, interacting with a single photon (red waveform) present between two photonic cavity mirrors. (b) Schematic representation of the bands of a 1DTSC before (left) and after (right) placing it in a cavity in the high-frequency regime, where zero energy of one sector does not overlap with conduction or valence bands from another. Here, the 0th photon sector has nontrivial topology, while the 1st photon sector is trivial. 
    Parameters: $\mu=-2t, \ \Delta=\alpha=0.2t, \ B=0.3t$ and for the cavity (right) $\omega=6t$,  $\gamma=0.4t$. 
    }
    \label{fig:schem}
\end{figure}

At the same time, light-matter interaction has been generating a lot of attention, not only as a probe field for equilibrium material properties~\cite{GiannettiPumpProbe2016}, but also for offering additional control knobs to engineer `on-demand' 
properties~\cite{BasovNM2017,BaoNRP2022}, such as light-induced metal-insulator transition~\cite{ManfredScience1998}, photovoltaic effect~\cite{OkaPRB2009}, photo-thermoelectric effect~\cite{XuNL2010}, and superconductivity~\cite{FaustiScience2011}.
Additionally, non-equilibrium generation of topological states of matter using a time-periodic laser pulse has also generated substantial interest~\cite{OkaPRB2009,FloquetGuPRL2011,UsajPRB2014,Eckardt2017,oka2019,BaoNRP2022,GhoshJPCMReview2024}. 
However, the coupling between matter and light is usually small (of the order of the fine structure constant in vacuum)~\cite{SchlawinAPR2022}, but it can be engineered to become strong in a photonic cavity, even in the presence of a few photons, allowing for quantum effects in the light-matter interaction~\cite{RitschRMP2013,SchlawinAPR2022}.
Accordingly, the use of quantum light to manipulate and engineer many-body properties of a quantum material represents a promising and significant research direction~\cite{RitschRMP2013,SchlawinAPR2022}. 
In particular, the experimental advent of high-quality light cavities~\cite{RaimondRMP2001,LeibfriedRMP2003} has empowered engineering of exotic material properties, such as supersolids~\cite{LeonardNature2017,LeonardScience2017}, charge transport~\cite{DavidPRL2017}, superconductivity~\cite{SchlawinPRL2019,CurtisPRL2019,KozinPRB2025}, and topology~\cite{WangXPRB2019,GuerciPRL2020,ChiocchettaNC2021,NguyenPRB2024,BacciconiPRX2025, JiangPRB2025, yangCommPhys2025, CardosoArXiV2025, YangArxiV2025, weiArxiV2025}. 

Another, extremely interesting avenue is to combine 1DTSC with the light-matter interaction of a photonic cavity.
From the theoretical perspective, cavity light-matter interaction problems have previously been solved using a mean-field ansatz~\cite{KozinPRB2025} and high-frequency approximation~\cite{JiajunLiPRL2020,SentefPRR2020,JiajunLiPRB2022}.
Moreover, in terms of topological states of matter, most focus has been on how topology affects the photonic degrees of freedom ~\cite{TrifPRL2012,SchmidtNJP2013,DmytrukPRB2015,DartiailhPRL2017,ContaminNQI2021,DmytrukPRB2023,Dmytruk.Marco.24}.
However, cavity effects on the robustness of topological edge states are still being discussed~\cite{BacciconiPRB2024,PerezGonzalez2025lightmatter}, while the poor man's MBS in a cavity has recently been studied from the perspective of the energy spectrum \cite{gomez_dmytruk_25}, but then topology still remains to be considered. 

In this work, we investigate a 1DTSC in a photonic cavity, see Fig.~\ref{fig:schem}(a) for a schematic of the system, both focusing on the properties of the MBS and establishing a robust and efficient method to characterize the topology of the combined light-matter system.
As a result, our work combines cavity quantum electrodynamics and topological superconductivity within a fully quantum description, where the number of photons, their coupling to electronic degrees of freedom, and the electronic degrees of freedom themselves in the case of a 1DTSC are all explicitly considered.
Due to the coupling of the 1DTSC to a cavity, we observe significant effects for the MBS.
Notably, the photon energy $\hbar \omega$ shifts the entire energy spectrum to higher energies, while the light-matter coupling induces a parameter dependence to the MBS energy, akin to a \textit{pseudo-dispersion}, in which the MBS energy varies with both light-matter coupling and magnetic field. This leads to a distinct tunability of cavity MBS. 
Furthermore, the cavity introduces an additional stabilizing effect on the MBS by significantly reducing the MBS oscillations, proportionally to the light matter coupling. We further find that disorder stability remains within the cavity, despite a somewhat smaller topological gap.

As a second main result, we characterize the topological phases emerging in the combined 1DTSC-cavity system.
We achieve this by employing the spectral localizer~\cite{loring.15}, a real-space and energy-resolved topological invariant, which provides a reliable and computationally effective method to assess topology at different energies as needed in a photonic cavity.
More precisely, for $N_{ph}$ photons in the cavity, we have in total $N_{ph}+1$ photon sectors, labeled by $N^{ps}$. Each photon sector contain a copy of the energy spectrum of the bare 1DTSC centered around energy $\omega(N^{ps}+\frac{1}{2})$ and modified by the cavity's light-matter coupling $\gamma$. See Fig.~\ref{fig:schem}(b) for a visualization when the cavity contains one photon.
In the high-frequency regime, we have multiple, well-separated photon sectors such that the MBS energies of one sector never overlaps with states from any other sector. In this regime we find notable light-induced changes to the topological phase diagram, all well-characterized by a standard spectral localizer.
However, in the low-frequency regime, when photon energy $\hbar\omega \lesssim W/2$, where $W$ is the bandwidth of the system, the MBS of one photon sector start to hybridize with bulk states from other photon sectors.
This raises a fundamental question about survival of the topological phase.
Unfortunately, standard spectral localizer formalisms here produce \textit{polluted} topological phase diagrams, with no connection to the actual topology of the system. We resolve this by judiciously engineering the energy term of the spectral localizer, which generates a clearcut and computationally effective identification of the topological state of a 1DTSC in a light cavity.
This allows for extending the scope of topological analysis in light-matter coupled hybrid systems beyond existing approaches.

The remainder of this work is organized as follows. In Section~\ref{Sec:Methods}, we describe the studied 1DTSC and how it couples to light in the photonic cavity, as well the scheme in which we use the spectral localizer, yielding a robust topological indicator.
In Section~\ref{sec:results}, we describe the results focusing on the regimes of high and low photon frequency where we analyze both the system's energy structure and topology, as well as sensitivity to disorder.
Finally, we summarize our work in Section~\ref{sec:summary}.

\section{Setup and methods} \label{Sec:Methods}

We start by presenting the essential theoretical details of the work, including the Hamiltonian for the 1DTSC, the photonic cavity inducing the light-matter coupling, and the spectral localizer, used to characterize the topology of the system.

\subsection{1D topological superconductor}

The electronic system studied in this work is described by the Hamiltonian $H=H_{\rm NW}+H_{\Delta}$.
The semiconductor nanowire with Rashba SOC is encapsulated by the Hamiltonian~\cite{Oreg2010,LutchynPRL2010,Leijnse_2012,Alicea_2012,Mourik2012Science,das2012zero,ramonaquado2017}
\begin{align}\label{eq:NormalH}
 H_{\rm NW} =&  \sum_{j, \sigma} \left( \mu + B\sigma_{\sigma\sigma}^z \right) c_{j \sigma}^\dagger c_{j  \sigma}^{} +\sum_{ j  \sigma}   t  c_{j \sigma}^\dagger c_{j+1 \sigma}  \non \\
 &+i \sum_{ j  \sigma \sigma^\prime} \alpha c_{j \sigma}^\dagger  \sigma_{\sigma\sigma^\prime}^y  c_{j+1  \sigma'} + {\rm H.c.} 
,\end{align}
where the operator $c_{j \sigma}^\dagger~(c_{j \sigma})$ creates~(destroys) an electron of spin $\sigma$ at site $j$ in the nanowire.
Here $\mu$ is the overall chemical potential, $B$ models an effective Zeeman splitting induced by an external magnetic field applied along the nanowire, $t$ is the hopping amplitude between nearest neighbor sites, and $\alpha$ is the strength of the Rashba spin-orbit coupling (SOC). The Pauli matrices $\vect{\sigma}$ act in the spin sector. 
We use $t$ as a unit of energy and set the length of the nanowire to be $100$ sites to decrease the MBS overlap between ends of the nanowire.
The nanowire is placed on top of a conventional spin-singlet bulk $s$-wave superconducting substrate.
Due to the proximity effect, an induced superconducting gap appears in the nanowire modeled by:
\begin{align}
 H_{\rm \Delta} = \sum_{j} \Delta c_{j  \uparrow}^\dagger  c^\dagger_{j \downarrow} + {\rm H.c.}
,\end{align}
where $\Delta$ is the (real-valued) proximity-induced superconducting gap. 

For an infinite, translation-invariant nanowire, we can write the Hamiltonian $H$ in momentum space using the Bogoliubov–de Gennes basis $\bf{\Psi}_{\textit{k}}$ = $\left\{ c_{k\uparrow}, c_{k\downarrow}, c^{\dagger}_{-k\uparrow},c^{\dagger}_{-k\downarrow} \right\}^T$ as 
\begin{align}
    H(k)=& (\mu + 2 t \cos ka) \ \tau_{z}\sigma_0 - 2 \alpha \sin ka \  \tau_{z} \sigma_{y} \non \\
    &+ B \ \tau_{z}\sigma_{z} +\Delta \ \tau_{y}\sigma_{y} ,
    \label{eq:Momentumhamiltonian}
\end{align} 
where the Pauli matrices $\vect{\tau}$ act in the particle-hole sector and $a$ is the lattice constant. The Hamiltonian $H(k)$ [Eq.~\eqref{eq:Momentumhamiltonian}] breaks time-reversal symmetry $\mathcal{T}^{-1} H(k) \mathcal{T} \neq H(-k)$, with $\mathcal{T}=i \tau_0 \sigma_y K$ and $K$ the complex-conjugation operator, but respects particle-hole symmetry $C^{-1} H(k) C = -H(-k)$ with $C=\tau_x \sigma_0 K$, and chiral symmetry $S H(k) S = -H(k)$ with $S=\tau_x \sigma_0$.
The Hamiltonian $H$ is topologically nontrivial, with MBS as topologically protected edge states, when $\left| B_{c_1} \right| < B < \left| B_{c_2} \right|$ with $\left| B_{c_1} \right| = \sqrt{(\mu + 2 t)^2 + \Delta^2}$ and $\left| B_{c_2} \right| = \sqrt{(\mu - 2 t)^2 + \Delta^2}$.

\subsection{Cavity light-matter interaction}
The 1DTSC is then placed in a photonic cavity. To understand the regime of a few photons, we model the interaction between the 1DTSC and a single photon mode of frequency $\omega$, which resonates with the cavity.
The cavity mode contributes to the total energy  with a term 
\begin{align}
 H_{\rm ph} =  \hbar \omega \left(b^{\dagger}b+\frac{1}{2}\right),
 \label{eq:Hbos}
\end{align}
where $b$ ($b^\dagger$) is the creation (annihilation) operator of the cavity light mode. The expression~\eqref{eq:Hbos} gives a frequency-dependent shift in the total energy, even in the absence of photons, due to zero-point energy induced fluctuations.

The coupling between light and matter in the cavity is handled through a Peierls substitution, now with bosonic operators for a full quantum description~\cite{svintsov.alymov.24,macedofaundez24},
\begin{align}
    \left(t,\alpha\right) 
    \rightarrow \left(t,\alpha\right) \exp\left[ \frac{i \gamma}{t} \left(b^{\dagger}+b\right)\right],
    \label{Eq:CavityPeierls}
\end{align}
where $\gamma = eatA_{\rm vac}/\hbar c$ is the coupling between light and matter, while the average amplitude of fluctuations is given by $A_{\rm vac} = \sqrt{\frac{2 \pi \hbar c^2}{\omega V}}$, with $V$ being the volume of the cavity.
Henceforth, we set the physical constants (electric charge, speed of light, and Planck constant, respectively) to unity $e=c=\hbar=1$. 
Thus, the 1DTSC placed in a cavity field is described as
\begin{align}\label{eq:RashbaHamPeierls}
 {\cal H}_{\rm C} =&  \sum_{j, \sigma} \left( \mu + B\sigma_{\sigma\sigma}^z \right) c_{j \sigma}^\dagger c_{j  \sigma}^{} +\sum_{ j  \sigma}   t e^{i\gamma \left(b^\dagger +b\right)/t}  c_{j \sigma}^\dagger c_{j+1 \sigma}  \non \\
 &+i \sum_{ j  \sigma \sigma^\prime} \alpha e^{i\gamma \left(b^\dagger +b\right)/t} c_{j \sigma}^\dagger  \sigma_{\sigma\sigma^\prime}^y  c_{j+1  \sigma'} + \sum_{j} \Delta c_{j  \uparrow}^\dagger  c^\dagger_{j \downarrow} + {\rm H.c.} \non \\
 &+ \omega \left(b^{\dagger}b+\frac{1}{2}\right) .
\end{align}

The quantum aspect of the coupling with the cavity is evidenced by the fact that, effectively, the fermionic Hamiltonian has multiple `copies' in different photon sectors, associated with different photon numbers, thereby forming an infinite-dimensional Hamiltonian
\begin{align}\label{Eq:InfiniteHam}
    \mathcal{H}_\infty=
    \begin{pmatrix}
         {\cal H}_{0,0} & {\cal H}_{0,1} & {\cal H}_{0,2} & \cdots  \\
         {\cal H}_{1,0} & {\cal H}_{1,1} & {\cal H}_{1,2} & \cdots  \\ 
         {\cal H}_{2,0} & {\cal H}_{2,1} & {\cal H}_{2,2} & \cdots  \\
         \vdots  & \vdots  & \vdots  &  \ddots  
    \end{pmatrix},
\end{align}
where ${\cal H}_{N,M}=\braket{N|{\cal H}_{\rm C}|M}$ represents the overlap matrix of the Hamiltonian $\cal H$ and the photon sectors with $N_{ph}=M$ and $N_{ph}=N$ photons. 
The bandwidth of each photon sector of $\mathcal{H}_\infty$ is $W \sim 8t$.
The diagonal blocks ${\cal H}_{M,M}$ result in a dressing of the electronic Hamiltonian in each different photon sector, while ${\cal H}_{N,M} ( \forall N \neq M)$ represents the coupling between different photon sectors. The latter creates a structure analogous to Floquet bands~\cite{OkaPRB2009,FloquetGuPRL2011,UsajPRB2014,Eckardt2017,oka2019,BaoNRP2022,GhoshJPCMReview2024}, but takes into consideration the full quantization of the photon degrees of freedom (and thus without the periodicity of the energy scale present in Floquet theory).
While presenting results, we focus on the case of finite number of photons, therefore in most cases we only retain the upper part of $\mathcal{H_{\infty}}$ matrix up to $N_{ph}$-th number of photons in the cavity, which we then denote as $\mathcal{H}_{N_{ph}}$.

Specifically, for the 1DTSC in a cavity we arrive at (see Appendix~\ref{app:matter_light} for details):
\begin{align}
{\cal H}_{N,M}=&\braket{N|{\cal H}_{\rm C}|M} \non \\
    =&\sum_{j,\sigma}  \left( \mu + B\sigma_{\sigma\sigma}^z \right) c^\dagger_{j\sigma}c_{j\sigma}+\sum_{j,\sigma} t^\gamma_{N, M} c^\dagger_{j\sigma} c_{j+1\sigma} \non \\
    &+i\sum_{ j  \sigma \sigma^\prime} \alpha^\gamma_{N, M}  \left[ \sigma_{\sigma\sigma^\prime}^y \right] c_{j \sigma}^\dagger c_{j+1  \sigma'} \sum_{j} 
    + \Delta c_{j  \uparrow}^\dagger  c^\dagger_{j \downarrow} \non \\
    & + {\rm H.c.} +\hbar \omega\left(N +\frac{1}{2}\right) \delta_{N, M} ,
    \label{Eq:RashbaHamwithlight}
\end{align}
where now the hopping amplitude $t^\gamma_{N,M}=t G_{N,M}(-\gamma)$ and the spin-orbit coupling $\alpha^\gamma_{N,M}=\alpha G_{N,M}(-\gamma)$  are the spin-conserving (and -flipping hopping, respectively, that incorporates the light-matter interaction ($\gamma$) in a given photon sector ($N=M$) or their coupling ($N \ne M$) and are thus matrices in the photon indices. 
Here, $G_{N,M}(x)$ is defined as
\begin{align}
    G_{N, M}(\gamma)=&\sum\limits_{n=0}^{N} \sum\limits_{m=0}^{M} e^{-\frac{\gamma^2}{2}}\frac{\left(-i \gamma \right)^{n+m}}{n!m!}\delta_{N-n, M-m}  \non \\
    &\times P(n, N)P(m, M) ,
    \label{eq:G}
\end{align}
where $P(a,b)$ is the square root of the Pochhammer symbol \begin{eqnarray}
    P(a, b)\equiv \sqrt{\frac{\Gamma(1+b)}{\Gamma(1+b-a)}} = \prod_{i=1}^{a} \sqrt{a-(b-i)},
    \label{eq:poch}
\end{eqnarray} with $\Gamma(x)$ being the Gamma function.

\subsection{Spectral localizer and topological invariant} \label{subsec:SpectralLocalizer}

Introducing different photon sectors enlarges the Hilbert space of the Hamiltonian, see $\mathcal{H_{\infty}}$ in Eq.~\eqref{Eq:InfiniteHam}, such that the system exhibits multiple copies of the original uncoupled Hamiltonian $H$ modified by the electron-photon interaction, transforming it into $\mathcal{H}_{NM}$. Figure~\ref{fig:schem}(b) depicts a schematic representation of the resulting bulk band structure. 
Since the multiple copies of the system are displaced in energy by $\omega$, the system can possibly host MBS at different energies $E=(N_{ph}+\frac{1}{2})\omega$. 
Thus, to topologically characterize MBS appearing at different energies, we require an energy-resolved topological invariant. 
To this end, we exploit the spectral localizer~\cite{loring.15,loring2017finitevolume,loring.19}. The spectral localizer (SL) is a local composite operator, consisting of the system's Hamiltonian and position operators as
\begin{align}
    L_{x,E}(X,\mathcal{H}_{\infty}) = \kappa \left(X-xI \right) \tau_x + \left(\mathcal{H}_{\infty} - E I \right) \tau_y,
    \label{Eq:SpectralLocalizer}
\end{align}
where $X$ represents the position operator, $x$ is the real space coordinate, and $I$ is the identity matrix. Here, the anti-commuting Pauli matrices $\vect{\tau}$ form the Clifford representation, and $\kappa$ is a (real-valued) coefficient ensuring a compatible weight between the Hamiltonian and the position operator. We set $\kappa=10^{-2}$ throughout this work unless mentioned otherwise. In Appendix~\ref{app:kappa}, we discuss the effect of $\kappa$ on the results, showing that the span of applicable values of $\kappa$ depends on the scheme used to extract a topological indicator from the SL.

One of the important quantities that can be extracted from the spectral localizer $L_{x,E}(X,\mathcal{H}_{\infty})$ is the localizer gap $\sigma(x,E)$, defined as the minimum absolute eigenvalue of $L_{x,E}(X,\mathcal{H}_{\infty})$~\cite{CerjanJMP2023,WongPRB2023}
\begin{align}
    \sigma(x,E)= {\rm min} \left( \lvert \sigma \left[L_{x,E}(X,\mathcal{H}_{\infty}) \right] \rvert \right),
    \label{eq:Lgap}
\end{align}
where $\sigma \left[L_{x,E}(X,\mathcal{H}_{\infty}) \right]$ is the spectrum of $L_{x,E}(X,\mathcal{H}_{\infty})$. 
The localizer gap $\sigma(x,E)$ is a local and energy-resolved quantity that vanishes if a topological boundary state exists at a given spatial location and energy, but remains finite elsewhere, including in the vacuum outside of the system, showing the topologically trivial nature of vacuum. Thus, the localizer gap $\sigma(x,E)$ acts like a local bandgap and can be employed to find the approximate location of topological edge states, also in driven systems~\cite{ghosh.arouca.24,ghoshMAC2025}.

We also extract the localizer index, which acts as a topological invariant. 
In this regard, we exploit the chiral symmetry $S$ of the system and define the topological invariant $\nu(x,E)$ as~\cite{loring.15,ghosh.arouca.24}
\begin{align}
     \nu(x,E)={\rm sig}([ \tilde{X}+i (\mathcal{H}_{\infty}-EI)]S) , 
     \label{Eq:localwind1D} 
\end{align}
where $\tilde{X}=\kappa (X-xI)$ and  ${\rm sig}$ represents the signature of a matrix, which counts the difference between the number of positive and negative eigenvalues. 
The topological invariant $\nu(x,E)$ is space-resolved and, importantly, also energy-resolved, which allows for probing topology under light-matter interactions despite the finite energies of the photon sectors. Notably, SL formalism requires us to only compute the eigenvalues of the matrix Eq.~\eqref{Eq:localwind1D}, making it a computationally efficient way to study topology of any system.

We note that the matrix $[\tilde{X}+i \mathcal{H}_{\infty}]S$ in Eq.~\eqref{Eq:localwind1D} is a Hermitian operator. However, the term $-iEIS$ makes the full expression in Eq.~\eqref{Eq:localwind1D} non-Hermitian. We resolve this by identifying the fact that the MBS will  appear at certain energies in the system, close to the photon energies $E=(N_{ph}+\frac{1}{2})\omega$, and we can therefore restrict the input energy to those particular energy values only~\cite{qi.na.24}. Afterward, we enforce the Hermiticity condition on $[\tilde{X}+i (\mathcal{H}_{\infty}-EI)]S$, i.e., we set the diagonal part to be real, such that the ${\rm sig}$-function is well-defined. We find generally that the topological invariant $\nu(x,E)$ takes a uniform value, depending on topological phase, inside the bulk of the system, while it vanishes near the boundary and outside of the system.  
To obtain phase diagrams using $\nu(x,E)$ we compute the average over the $20$ middle lattice sites of the nanowire, denoting it simply as $\nu$ while specifying the photon sector.
Finally, we report substantial problems in determining the topological phase using the SL without the chiral symmetry $S$, in the low-frequency regime.
We summarize this in Appendices~\ref{app:kappa} and \ref{app:sl}, where we compare the results for the SL with and without chiral symmetry $S$ from different perspectives.

\section{Results}
\label{sec:results}
We study a 1DTSC in a photonic cavity by solving Eq.~\eqref{Eq:RashbaHamwithlight} for energies and corresponding wavefunctions, and extract the topological invariant using the SL gap Eq.~\eqref{eq:Lgap} and invariant Eq.~\eqref{Eq:localwind1D}. We divide our results into two different regimes: the high-frequency regime with photon frequency larger than the system's half-bandwidth $\omega > W/2$, which allows a separation of the zero-energy level of one photon sector from all other photon sectors, and the low-frequency regime where such separation is not possible. 
But also before that, we note that even in the zero-photon case, i.e., when no photons are present in the system (\NPH{0}), which we refer to as the $0$th photon sector, given by ${\cal H}_{0,0}$ in Eq.~\eqref{Eq:InfiniteHam}, both the spin-conserving and spin-flipping hopping terms are still renormalized by the light-matter coupling $\gamma$ due to vacuum fluctuations in the cavity.
As a consequence, we find the conditions for the emergence of the MBS being modified as
\begin{align}
    B_{c_1}=&\sqrt{(\mu+2t)^2+\Delta^2} \rightarrow B_{c_1}^\prime=\sqrt{(\mu+2t^{\gamma}_{N, M})^2+\Delta^2}, \non \\
    B_{c_2}=&\sqrt{(\mu-2t)^2+\Delta^2} \rightarrow B_{c_2}^\prime=\sqrt{(\mu-2t^{\gamma}_{N, M})^2+\Delta^2}.
    \label{eq:conditiongamma}
\end{align}
For a finite number of photons, the situation becomes richer as we also observe transitions between different photon sectors that nontrivially affect the fermionic degrees of freedom.
In particular, we find that the presence of photons changes the condition for the emergence of MBS in a nontrivial manner, as reported below.

\subsection{Energy spectrum in the high-frequency regime}
\begin{figure}[thb]
    \centering
    \includegraphics[width=0.9\linewidth]{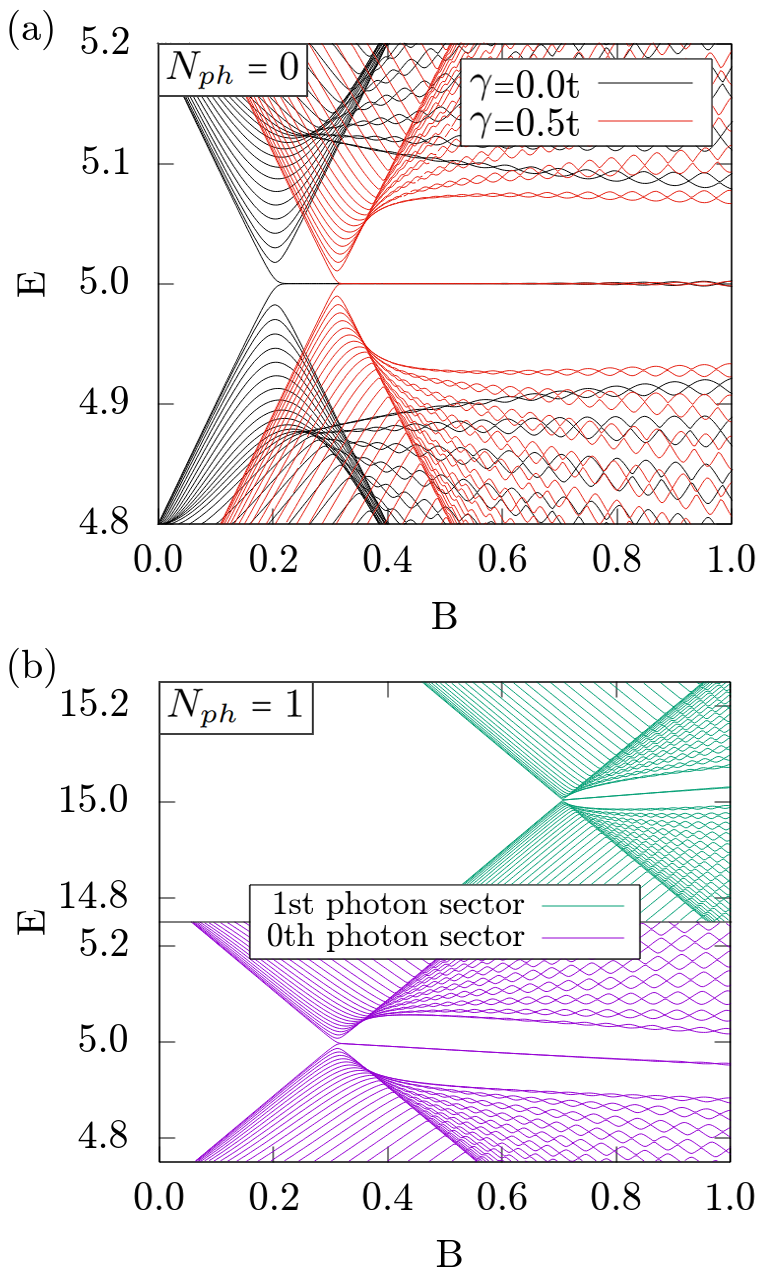}
    \caption{(a) Eigenvalues of $\mathcal{H}_{0}$ [Eq.~\eqref{Eq:InfiniteHam}] as a function of magnetic field $B$ for uncoupled ($\gamma=0.0$, black) and coupled ($\gamma=0.5t$, red) system in the 0th photon sector for $N_{ph}=0$. (b) Eigenvalues of $\mathcal{H}_{1}$ as a function of $B$ in the 0th (green) and 1st (violet) photon sector for $N_{ph}=1$. Bottom~(top) panel shows eigenvalues close to $\omega/2$~($3\omega/2$), corresponding to the bare photon energy of the $0$th~($1$st) photon sector. Parameters are $\mu=-2t, \ \Delta=\alpha=0.2t$, $\omega=10t$, and $\gamma=0.5t$ for (b).}
    \label{fig:eigens}
\end{figure}

We start by discussing the energy spectrum in the high-frequency regime, $\omega \gtrsim W/2$. We first focus on zero or one photon in the cavity and plot the low-energy spectrum in Fig.~\ref{fig:eigens}.
Even in the case of no photons, $N_{ph}=0$, in Fig.~\ref{fig:eigens}(a),  the direct influence of the cavity on the 1DTSC is clear due to the entire energy spectrum being shifted by $\frac{\omega}{2}=5t$ (black lines). As such, the MBS zero-energy modes appearing after the bulk gap closing are now located at $\frac{\omega}{2}=5t$.
Turning on a finite light-matter coupling $\gamma$ (red lines), the 1DTSC interacts with vacuum fluctuations in the cavity, which has an observable effect on the MBS energy spectrum.
In particular, we find that the topological gap size decreases in comparison to the non-coupled case. However, as we find in Section \ref{subsec:disorder}, this does not reduce disorder stability. Additionally, the critical magnetic field $B_{c_1}$ for entering the topological phase increases with $\gamma$, in agreement with Eq.~\eqref{eq:conditiongamma}, as $\gamma$ directly influences the values of both the spin-conserving and spin-flipping hopping amplitudes $t^\gamma$ and $\alpha^\gamma$. Sufficiently large light-matter coupling can even lead to the closing of the topological gap, showing the clear influence of the cavity on the topology of the system.

\begin{figure}[tbh]
    \centering
    \includegraphics[width=0.9\linewidth]{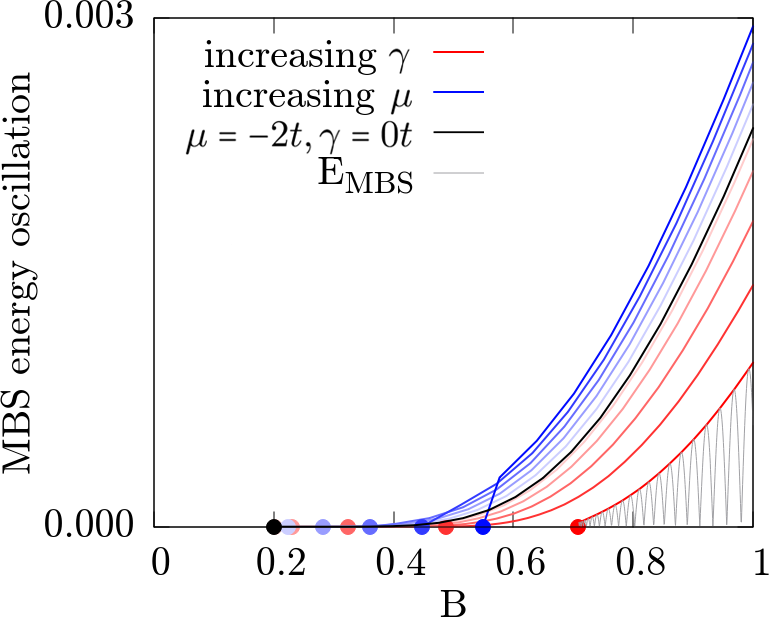}

    \caption{Envelope plot of the MBS energy oscillations for increasing chemical potential $\mu=-2t \rightarrow \mu=-1.5t$ in the 0th photon sector for the bare 1DTSC for $\gamma = 0$ (blue lines) and increasing light-matter coupling $\gamma=0t \rightarrow \gamma=0.5t$ of the 1st photon sector for $\mu = 2t$ (red lines).
    Same colored dots mark the value of the critical magnetic field $B_{c_1}$ for each value of the varied parameter. 
    Energy spectrum $E_{\rm MBS}$ (grey lines) shows the oscillations of a single MBS level enveloped by the line depicting the largest shift from the initial value ($\mu=-2t$ and $\gamma = 0.5t$). 
    Parameters are $B=0.4t, \ \Delta=\alpha =0.2t$,  $\omega = 10t$, $N_{ph}=1$.
    }
    \label{fig.envelope}
\end{figure}

\begin{figure*}
    \centering
    \includegraphics[width=0.9\linewidth]{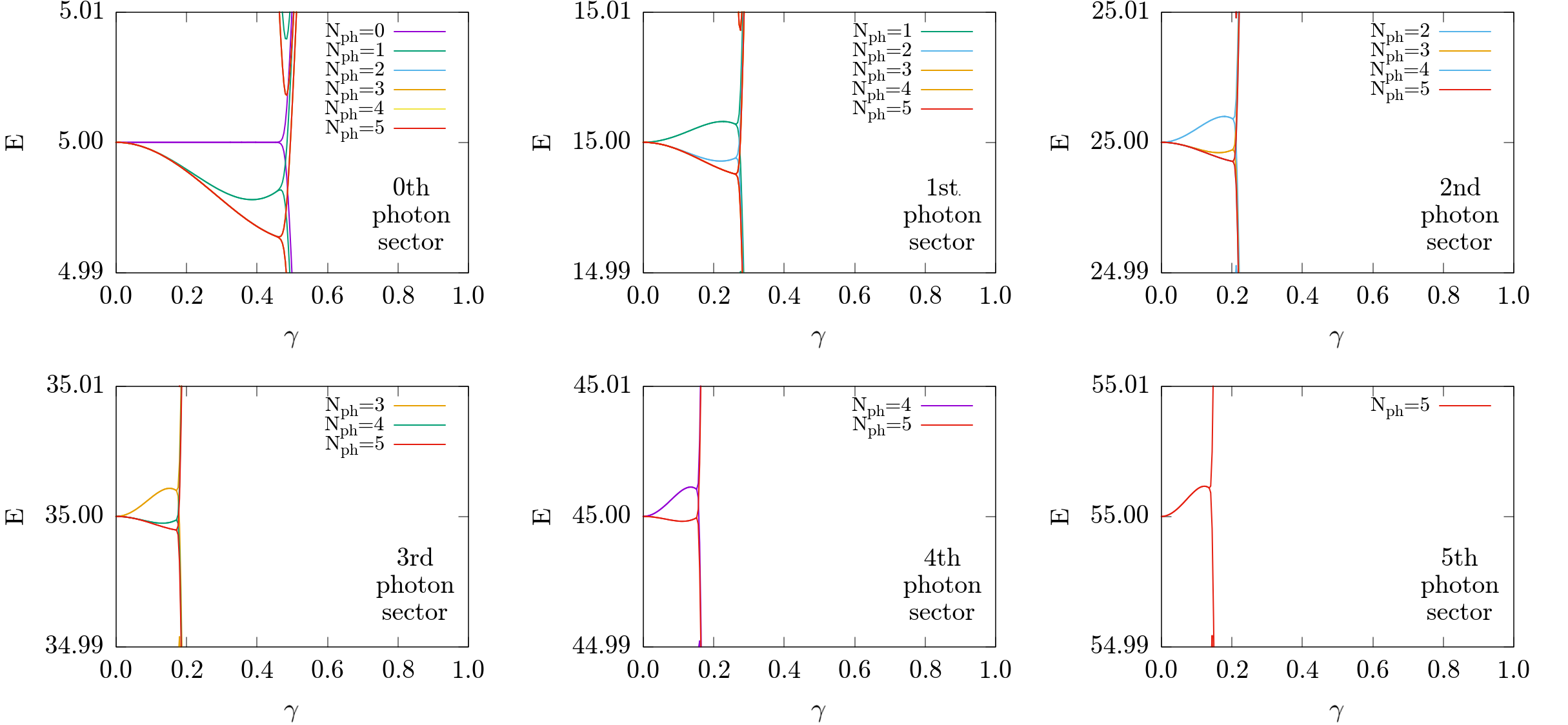}
    \caption{MBS energies of $\mathcal{H}_{0-5}$ for the lowest 6 photon sectors solved for up to $N_{ph} = 5$ photons. Apart from the case of $N_{ph}=0$, where the MBS do not exhibit an energy \textit{pseudo-dispersion} (PD), the MBS energy is each photon sector is either increased (highest photon sector for given $N_{ph}$), decreased (photon sectors $N_{ph}-1$ and $N_{ph}-2$) for finite $\gamma$. Parameters are $\mu=-2t, \ \Delta=\alpha=0.2t, \ N_{ph}=1, \ B=0.3t$ and $\omega=10t$.
    }
    \label{fig:Nph}
\end{figure*}

In Fig.~\ref{fig:eigens}(b), we show the energy spectrum in the case of one photon inside the cavity, \NPH{1}.
In this case, we have two photon sectors, $0$th (violet) and $1$st (green), centered around $\frac{\omega}{2}=5t$ and  $\frac{3\omega}{2}=15t$, respectively.
As seen, each sector has distinct values of $B_{c,1}$ for transitioning into the topological regime. In general, we find that higher order photon sectors require higher magnetic fields to reach the topological phase for nonzero $\gamma$.
Most interestingly, the MBS are now not tied to energies $\omega(N^{ps}+\frac{1}{2})$, where $N^{ps}$ is the photon sector, but contrary to well-established MBS behavior~\cite{Oreg2010,LutchynPRL2010}, the  MBS energy also becomes a function of the magnetic field $B$. 
We refer to this modulation of the MBS energy as a \textit{pseudo-dispersion} (PD), as it changes the energy level as a function of the magnetic field, while light-matter coupling $\gamma$ changes the slope of PD.
We attribute the PD to the renormalization of bare photon energy as the photon degree of freedom is incorporated into the fermionic degrees of freedom and to the existence of nonzero terms $\mathcal{H}_{NM}$ with $N \ne M$, mediating the coupling between separate photon sectors. Due to the latter, the MBS does not exhibit a PD for $N_{ph}=0$.

The two sets of overlying MBS energy levels in Fig.~\ref{fig:eigens}(a) hint at a subtle yet significant effect of cavity-induced stabilization of MBS: 
Despite the topological gap protecting the MBS being smaller for finite $\gamma$, the MBS still experiences smaller energy oscillations. We investigate this effect in Fig.~\ref{fig.envelope} by extracting the envelope tracing the energy maxima of the MBS oscillations as a function of the magnetic field for both changing $\mu$ and $\gamma$. 
Starting from no light-matter coupling $\gamma=0$ and the optimal 1DTSC configuration at $\mu =-2t$, which has the smallest $B_{c,1}$ (black line), we see that increasing $\mu$ (blue lines) in the 0th sector increases the energy oscillations and increases $B_{c,1}$. Surprisingly, increasing $\gamma$ in any photon sector instead leads to diminishing oscillations, along with a larger $B_{c,1}$. Hence, the cavity acts as a stabilizer for the MBS, as it negates the MBS oscillations known to stem from various sources, e.g., high magnetic fields or finite size effects \cite{Li_16,theiler_19}. 
Notably, in the 0th photon sector, the stabilizing effect is also present, but its magnitude is less pronounced than in the 1st photon sector.
We find a similar suppression of MBS energy oscillations in all photon sectors.

We next investigate the influence of the number of photons on the 1DTSC by analyzing the MBS energy in different photon sectors in Fig.~\ref{fig:Nph}, with up to \NPH{5} photons present in the cavity, but the conclusions are general for all numbers of photons.
Here, we show a close-up of the MBS energy levels as a function of the light-matter coupling constant $\gamma$, choosing a magnetic field such that we are in the topological phase for $\gamma =0$.
An energy spectrum exists only in a specific photon sector, when the probed photon sector is equal to or smaller than the number of photons in the system, i.e.~ $N^{ps} \leq N_{ph}$, as each photon sector represents a replica of the system with its energy shifted by the photon energy $\omega$ from each other. 
Thus, as we plot photon sectors 0th to 5th in Fig.~\ref{fig:Nph}, the energy increases with $\omega$, and there are fewer and fewer energy spectra present in each panel.
Starting with the 0th photon sector, we observe for  $N_{ph} = 0$ photons in the cavity that the MBS have a constant energy as a function of the light-matter coupling $\gamma$.
For every other case, the MBS exhibit a PD, here as a function of $\gamma$.
\begin{figure*}
    \centering
    \includegraphics[width=0.30\linewidth]{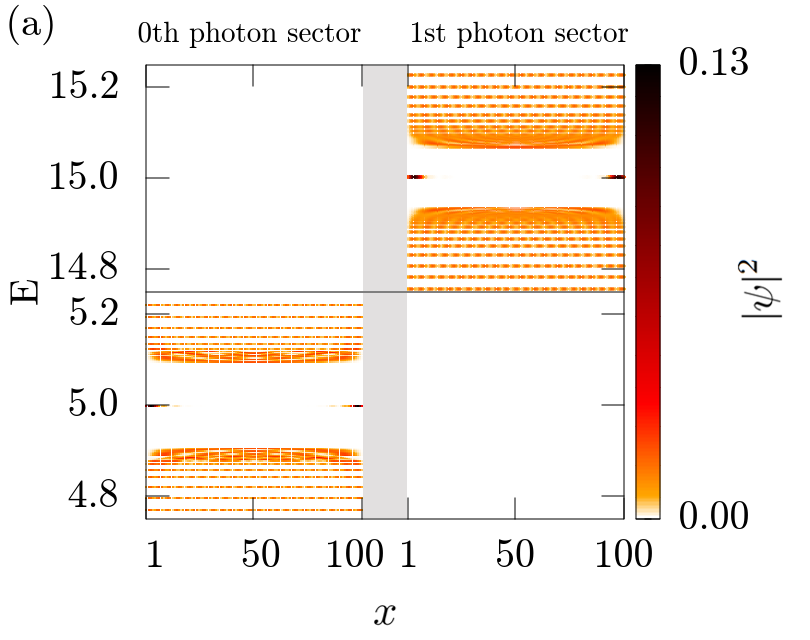}
    \includegraphics[width=0.30\linewidth]{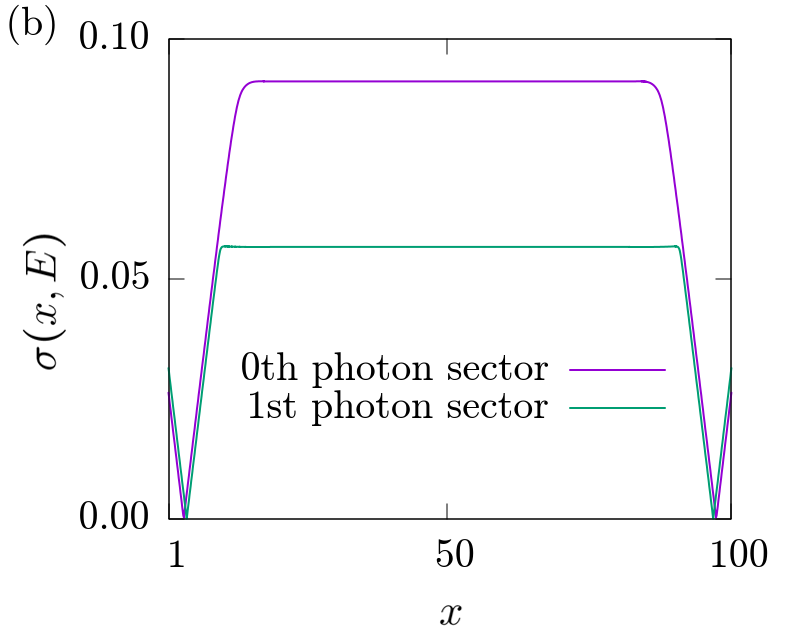}
    \includegraphics[width=0.30\linewidth]{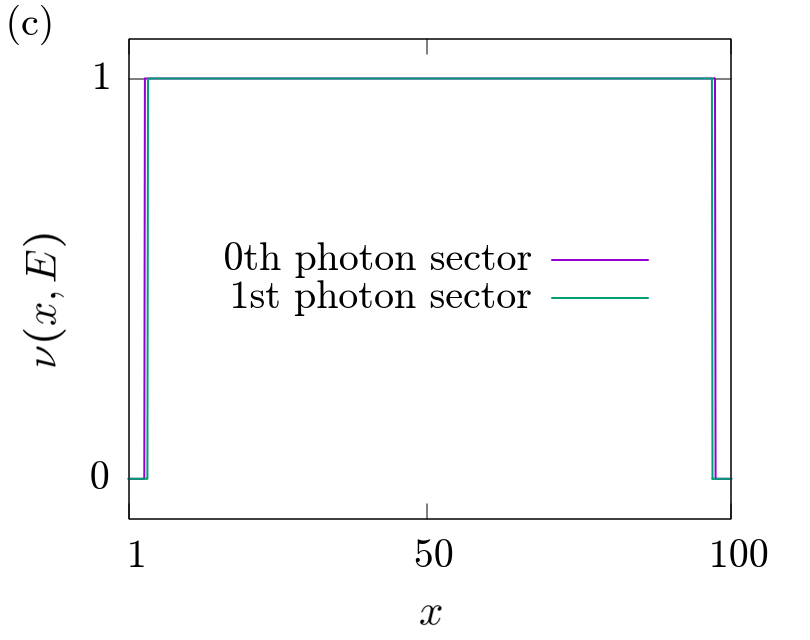}
    \caption{(a) Eigenvector distribution of $\mathcal{H}_1$ showing the MBS localization at the ends of the nanowire for both photon sectors. (b) Localizer gap $\sigma(x,E)$ and (c) topological invariant $\nu(x,E)$ for both photon sectors showing the existence of a topological phase in the system (i.e.~$\nu(x,E)=1$), with borders coinciding with the position of MBS in the nanowire.
    Parameters are  $\mu=-2t, \ \Delta=\alpha=\gamma=0.2t, \ N_{ph}=1, \ B=0.3t $ and (b,c) $\omega=10t, E=\omega/2$ ($0$th photon sector) and $E=3\omega/2$ ($1$st photon sector). }
    \label{fig:topo1}
\end{figure*}
For the PD, a certain regularity can be observed: when we probe the highest photon sector for a given number of photons $N_{ph}$, the PD results in the MBS energy always obtaining a extra positive energy contribution.
In contrast, for the rest of the photon sectors, the PD results in the MBS energy levels decreasing in energy.
Moreover, when the number of photons is $N_{ph}>N^{ps}+2$ in the $N^{ps}$th sector, the energy spectrum is degenerate, such that it is sufficient to solve for three photon numbers in each sector.
It is important to note that this energy modulation stemming from the PD is completely unrelated to finite size effects. We also note that higher photon sectors see the MBS disappear at an increasingly smaller $\gamma$, indicating a loss of non-trivial topology with increasing light-matter coupling.

To summarize, we observe that the energy spectrum of a 1DTSC is changed both qualitatively and quantitatively when the 1DTSC is placed in a photonic cavity.
Particularly, the MBS energy develops a pseudo-dispersion (PD), with the energy changing as a function of physical parameters, such as light-matter coupling $\gamma$ and magnetic field $B$. Moreover, in each photon sector, three different MBS energy levels appear due to coupling to different numbers of photons, all with different PDs.
This implies that we can manipulate the energy of the MBS with $\gamma$ and then further strengthen the effect by tweaking the magnetic field $B$. 
The cavity can further be used to stabilize the MBS by negating the MBS oscillations by increasing $\gamma$. Notably, this is despite a decreasing energy gap.

\subsection{Topology in the high-frequency regime}
The  analysis of the energy spectrum of MBS above is not sufficient to give a clear answer about topology. In this subsection, we explicitly verify the existence of nontrivial topology in a 1DTSC placed in a photonic cavity.
Due to the emergence of the MBS at $E \ne 0$ and the clear PD further changing the energy, the topological indicator must be energy-resolved.
We therefore choose the spectral localizer index $\nu(x,E)$ as a topological invariant, along with the localizer gap $\sigma(x,E)$, to define the topology of the $N_{ph}+1$ photon sectors, or replications of the system, for each fixed number of photons $N_{ph}$.
It is important to note that, despite the MBS changing their energy due to the PD, the spectral localizer in Eq.~\eqref{Eq:localwind1D} does not require any precise energy fine-tuning to be able to detect topological phases.
Instead, we can employ energies in a sufficiently wide range around the bare energies $(N_{ph}+\tfrac{1}{2})\omega$ and still identify the correct topological phase.

We begin the analysis of topology in the high-frequency regime ($\omega > W/2$) and for $N_{ph}=1$ in Fig.~\ref{fig:topo1}.
First, it is essential to identify the spatial features of the MBS, which are a pair of in-gap states localized at the edges of the system.
In Fig.~\ref{fig:topo1}(a), we therefore show the site- and energy-resolved eigenvector distribution, $|\psi|^2$, as a function of position $x$ along the nanowire, around $E=\omega/2$ (0th photon sector) and $E=3\omega/2$ (1st photon sector). 
We observe states, identified as the MBS, clearly localized towards the end points of the nanowire in both photon sectors.
Stemming from this, we know where the MBS are in both real space ($x$) and  energy ($E$) and thus, we know the full set of parameters needed for verifying the existence of non-trivial topology. We illustrate this by calculating the localizer gap $\sigma(x,E)$ and localizer index $\nu(x,E)$ in Figs.~\ref{fig:topo1}(b,c) as a function of position. 
As seen, the localizer gap closes ($\sigma(x,E)=0$) near the ends of the nanowire, exactly where the MBS is localized, while it remains finite in the bulk and outside of the system in both photon sectors. 
Note that the gap closing does not occur at the very edges of the nanowire but slightly into the nanowire because of finite size effects smearing MBS over a small section near the end of the nanowire, as also evident in Figs.~\ref{fig:topo1}(a).
Moreover, the localizer index $\nu(x,E)$ is finite inside the bulk of the system, but vanishes near and outside of the nanowire boundary.
Thus both localizer index $\nu(x,E)$ and the localizer gap $\sigma(x,E)$ in a one-to-one way indicate non-trivial topology in the nanowire, with MBS being the accompanied topologically protected boundary states.

\begin{figure}
    \centering
    \includegraphics[width=\linewidth]{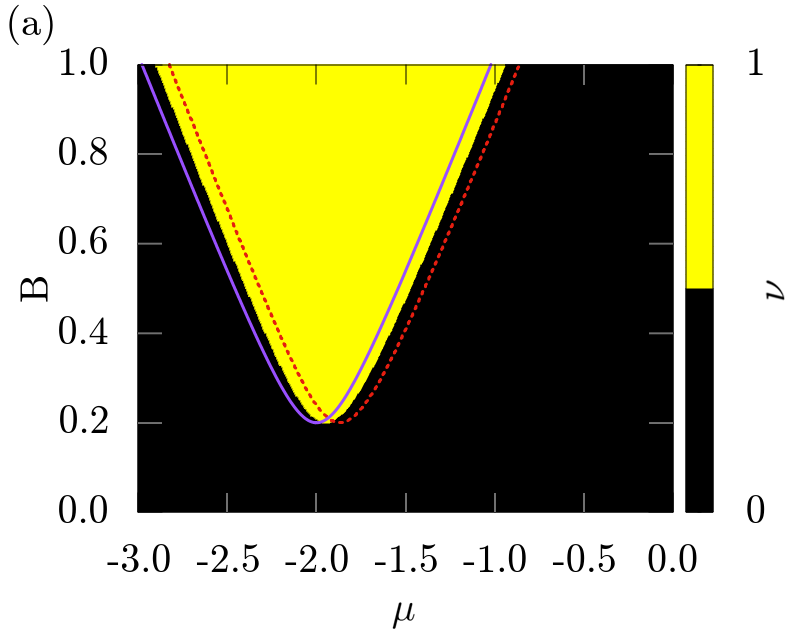}\\
    \includegraphics[width=\linewidth]{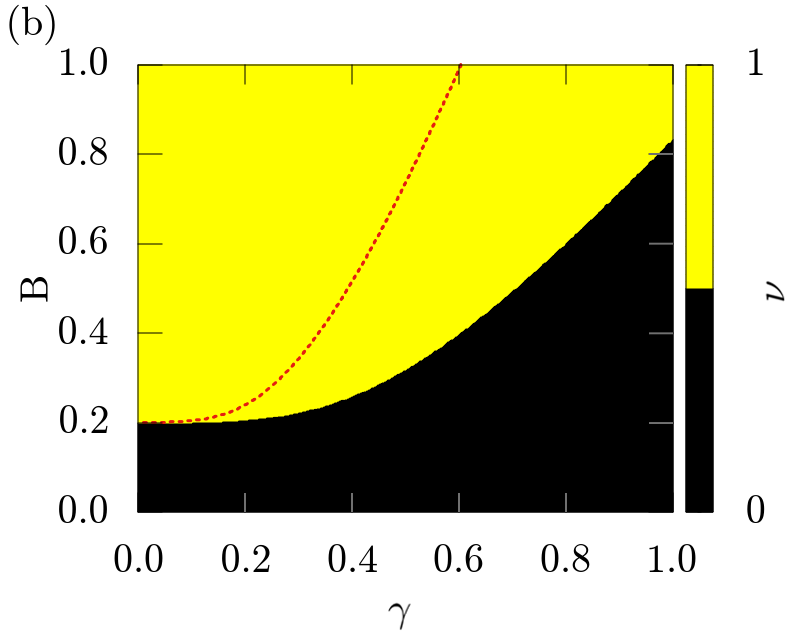}
    \caption{(a) Topological phase diagrams for (a) $\mu$ vs.~$B$ and (b)  $\gamma$ vs.~$B$ for $0$th photon sector ($E=\omega/2$). Red dashed line marks the edges of the topological phase for $1$st photon sector ($E=3\omega/2$) and violet solid line corresponds to the bare 1DTSC case (\NPH{\omega=\gamma=0}).
    Parameters are, unless indicated in figure, $\mu=-2t, \ \Delta=\alpha=\gamma=0.2t, \ \omega=10t, \ N_{ph}=1$.}
    \label{fig:topo2}
\end{figure}

Having topologically characterized the MBS using an energy- and space-resolved topological invariant, we proceed to investigate the topological phase diagrams.
In Fig.~\ref{fig:topo2}, we show the topological invariant $\nu$ in the $\mu \mhyphen B$ and $\gamma \mhyphen B$ planes for $N_{ph}=1$ in the $0$th photon sector, where the yellow (black) colored region is the topologically non-trivial (trivial) phase. 
In Fig.~\ref{fig:topo2}(a), displaying the $\mu \mhyphen B$ phase diagram for finite $\gamma$, the well-known topological phase transition line of the bare ($\gamma=0$) 1DTSC~\cite{Zhang2017NatCommun} is also plotted (violet line).
Here, we observe a shift between the yellow region and the violet curve, a deviation induced by the light-matter interaction $\gamma$.
Notably, this manifests even without photons present in the cavity for nonzero $\gamma$, due to vacuum fluctuations inducing an effective change of the hopping $t$, requiring the chemical potential $\mu$ to decrease to still reach the topological phase at fixed magnetic field $B$. 
We find that the shape of the yellow region in Fig.~\ref{fig:topo2}(a) can be reproduced by solving Eq.~(\ref{eq:conditiongamma}).
Moreover, we also plot the boundaries of the topological phase for the $1$st photon sector (red dashed line). This shows an additional shift between the different photon sectors.

In Fig.~\ref{fig:topo2}(b), we show the topological invariant $\nu$ in the $\gamma \mhyphen B$ plane for fixed chemical potential $\mu$. This topological phase diagram is unique to the light-matter coupled system.
It shows a clear change of the critical magnetic field $B_{c_1}$ due to $\gamma$, which becomes even more prominent in the $1$st photon sector (red-dashed line). 
There is thus a clear mismatch between topological phases for the two different photon sectors from the perspective of the shift in critical magnetic field $B_{c_1}$.
We expect this trend to continue for higher photon sectors stemming from our analysis in Fig.~\ref{fig:Nph}, with strong light–matter coupling ultimately becoming increasingly detrimental to topology as the photon sector increases, or equivalently, at higher energies.
Nevertheless, Fig.~\ref{fig:topo2}(b) reveals that the MBS survives for an extensive range of the light-matter coupling $\gamma$.
From an experimental point of view~\cite{anappara09,scalari12,maissen14,Gambino2014,Chikkaraddy2016,Yoshihara2017,flick17,Bayer2017,genco17,FriskKockum2019,Halbhuber2020}, the range is $\frac{\gamma}{\omega} \sim 0.1-1$ depending on the materials used, which clearly fall within the studied parameter regime.

\subsection{Effects of disorder}
\label{subsec:disorder}
As already seen in Fig.~\ref{fig:eigens}, the cavity light-matter interaction affects the size of the topological gap. 
It is thus interesting to investigate if this decrease has any consequence on the stability of the topological phase as the gap is notably protecting the MBS from disorder effects. 
To this end, we add on-site random Anderson-type disorder in the chemical potential to the Hamiltonian Eq.~\eqref{eq:RashbaHamPeierls}, given as $\sum_{j,\sigma} c^\dagger_{j\sigma} V_j c_{j\sigma}$. Here, $V_j$ is uniformly distributed in the interval $[ -D, D]$, with $D$ denoting the maximum disorder strength. The addition of disorder in the 1DTSC takes the modeling closer to the experimental setup, where disorder is unavoidable. We compute the topological invariant $\nu$ for each disorder sample and then take an average over $250$ random disorder configurations to obtain a disorder-averaged topological invariant $\bar{\nu}$. 

\begin{figure}
    \centering
    \includegraphics[width=1\linewidth]{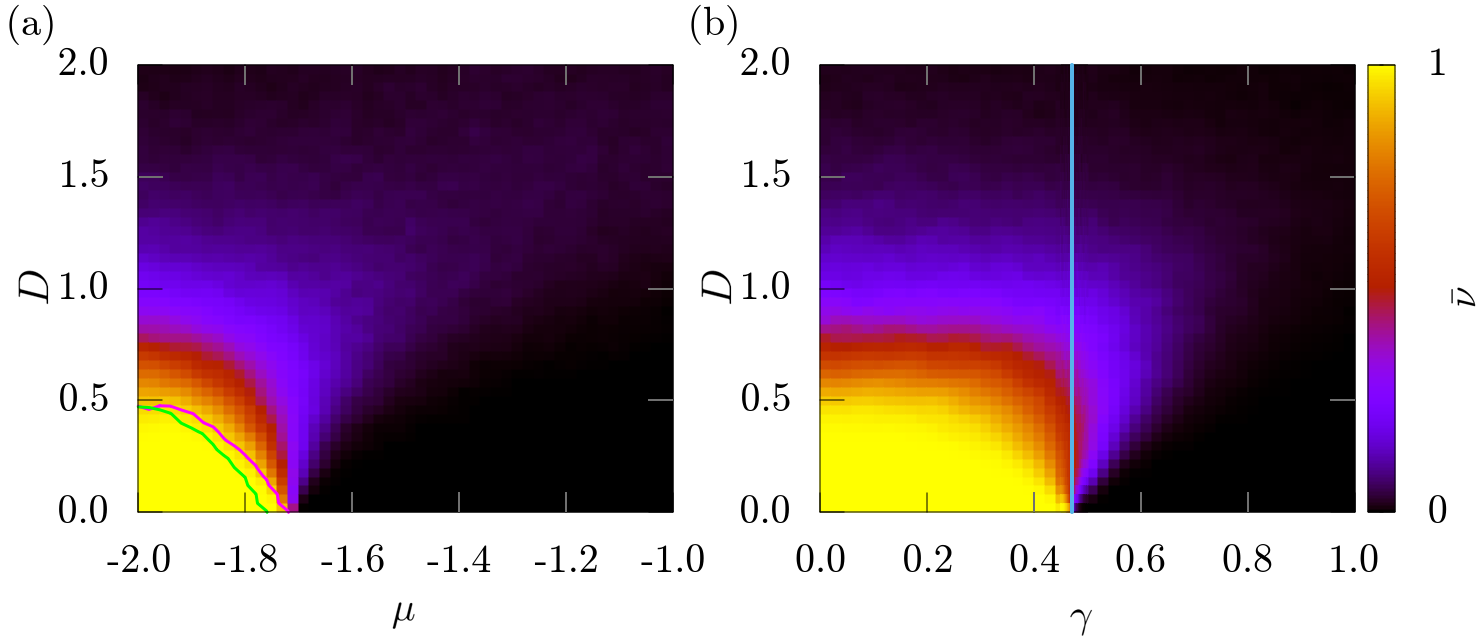}
    \caption{(a) Topological phase diagrams in the 0th photon sector (a) as a function of chemical potential $\mu$ vs.~disorder strength $D$.
    Green and magenta lines bound the regions with $\nu \simeq 1.0$ for $\gamma=0t$ (green) and $\gamma=0.2t$ (magenta). (b) Topological phase diagrams as a function of light-matter coupling $\gamma$ vs.~$D$. 
    Cyan line marks the topological transition at $\gamma\simeq 0.47t$ (see topological transition in Fig.~\ref{fig:topo2}(b) at $B=0.3t$).
    Parameters are  $\mu=-2t, \ \Delta=\alpha=0.2t, \ B=0.3t,\ \omega=10t, \ N_{ph}=1$, \ $E=\omega/2$ with (a) $\gamma=0.2t$ and (b) $\mu=-2t$. Results are obtained by averaging over $250$ disorder realizations.
    } 
    \label{fig:disorder}
\end{figure}

In Fig.~\ref{fig:disorder}(a), we show the dependence of $\bar{\nu}$ on disorder $D$ and chemical potential $\mu$ for finite $\gamma$ and $N_{ph} =1$ in the 0th photon sector.
The dome-like structure of the quantized topological phase is a result of the random disorder potential inducing impurities, which eventually fill the superconducting gap and thus no topological phase can develop for large enough $D$.
However, this phenomenon is not due to the cavity, as we find the same destruction of the topological phase by disorder even in a bare 1DTSC. This is illustrated by the similarity between the green and magenta lines, which bound the regions with $\nu \simeq 1.0$ for bare ($\gamma = 0t$) and light-coupled ($\gamma = 0.2t$) 1DTSC, respectively. 
The results in the bare limit are in agreement with previous works~\cite{akhmerov11,sau13,awoga17,kobialka20}. Notably, the cavity even allows for a somewhat larger disorder robustness at non-optimal $\mu>-2t$.
Therefore, non-zero light-matter coupling in a cavity has no detrimental effects on the topological phase robustness in comparison to the bare 1DTSC.

In Fig.~\ref{fig:disorder}(b), we show the dependence of the topological indicator $\bar{\nu}$ on the disorder $D$ and light-matter coupling $\gamma$.
The cyan line marks the topological phase boundary for the clean case. As seen, the topological region develops again a dome-like shape, suggesting that the topological phase gets destroyed by a strong disorder $D$. 
Nevertheless, when the system is deep in the topological phase, the robustness to disorder is independent of the value of $\gamma$ and thus, of the cavity.
The diminished robustness near the topological transition point (cyan line) manifests only because the size of the topological gap tends to zero, and thus, even small disorder is destructive.
Results for the 1st photon sector are quantitatively comparable for both the $\mu \mhyphen D$  and $\gamma \mhyphen D$ cases. 
To summarize, despite the smaller topological gap for the 1DTSC in a cavity, the stability of the topological phase is not affected in any significant way. A such, a photonic cavity offers an additional tunability without any notable drawbacks.

\subsection{Energy spectrum and topology in low-frequency regime}
\label{subsec:lowfreq}
The high-frequency region in the previous subsections has all photon sectors clearly separated.
In this subsection, we discuss the regime of low-frequency photons ($\omega \lesssim W/2$), where energy states from different photon sectors begin to significantly overlap with each other.
Here, as previously, we consider the case of \NPH{1}; however, the conclusions also follow if the cavity contains a higher number of photons. 

\begin{figure}
    \centering
        \includegraphics[width=0.8\linewidth]{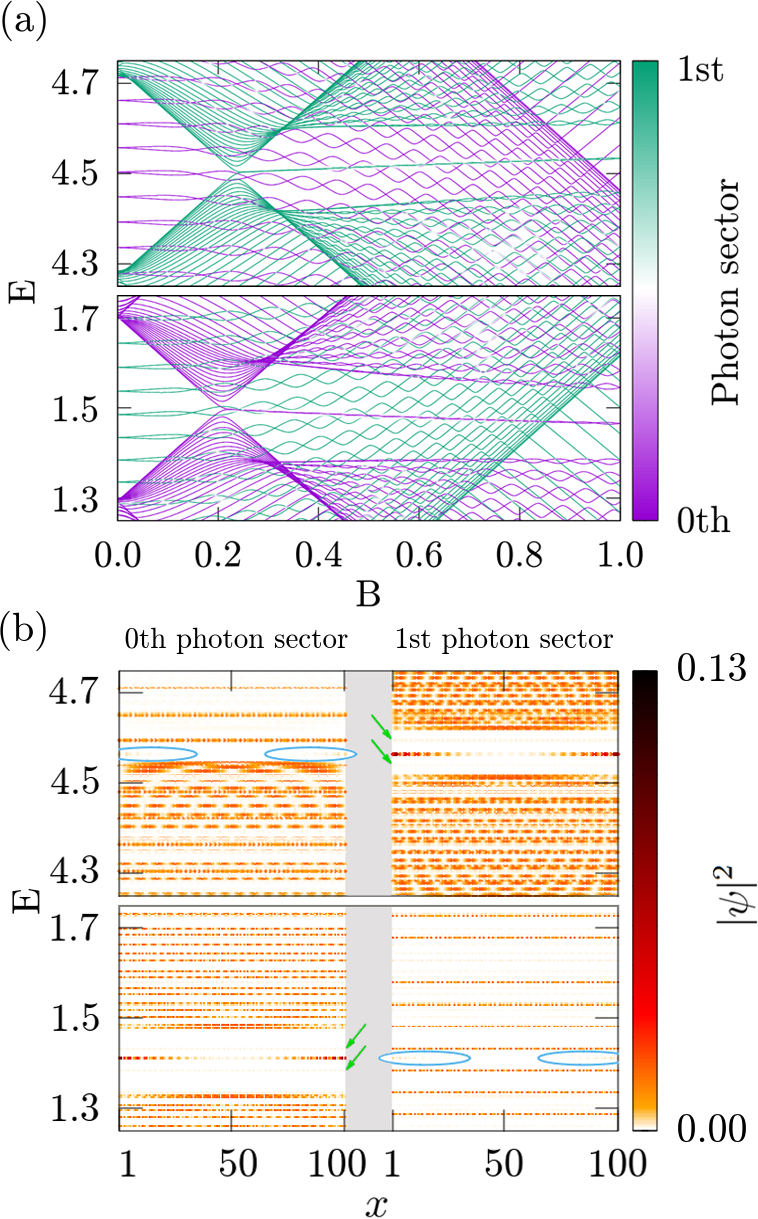}
    \caption{(a) Eigenvalues of $\mathcal{H}_1$ [Eq.~\eqref{Eq:InfiniteHam}] as a function of magnetic field $B$ in the 0th (green) and 1th (violet) photon sector for $N_{ph} =1$. 
    Bottom (top) panel shows eigenvalues close to $\tfrac{1}{2}\omega$ ($\tfrac{3}{2}\omega$), corresponding to the
    bare photon energy of the 0th (1st) photon sector. 
    (b) Eigenvector distribution of $\mathcal{H}_1$ [Eq.~\eqref{Eq:InfiniteHam}] showing the MBS localization at the ends of the nanowire for both photon sectors. Blue ovals (green arrows) indicate leakage of MBS (of bulk states) into the opposite photon sector.
    Parameters are $\mu=-2t, \ \Delta=\alpha=0.2t, \ \gamma=0.2t, \ \omega=3t, \ N_{ph}=1$, and (b) $\gamma=0.4t$ and $B=0.8t$.}
    \label{fig:topo3}
\end{figure}

We start by investigating the energy spectrum around $E=\omega/2$ and $E=3\omega/2$ as a function of the magnetic field $B$ in Fig.~\ref{fig:topo3}(a) for the 0th (violet) and 1st (green) photon sectors. In contrast to Fig.~\ref{fig:eigens}(b) in the high-frequency regime, energies from the two photon sectors are now overlapping.
While the MBS and its energy dependence is still preserved and exist around the bare energies, $\omega/2$ and $3\omega/2$, including a PD as found earlier, the entire topological gaps are now filled by states from the other photon sector. 
In Fig.~\ref{fig:topo3}(b), we present the energy- and space-resolved eigenvector distribution $|\psi|^2$ along the nanowire, in the same energy regime as in (a). Beyond the MBS, all other states extend throughout the nanowire as bulk states.
We also observe signatures of hybridization between photon sectors:
The MBS in one photon sector leaks into the other photon sector (blue ovals) and the topological gap around the MBS hosts states from the other photon sector (green arrows). We also find such hybridization in the bulk continuum.
This shows that states from different photon sectors overlapping in energy are also actively hybridizing with each other.
\begin{figure}
    \centering
        \includegraphics[width=\linewidth]{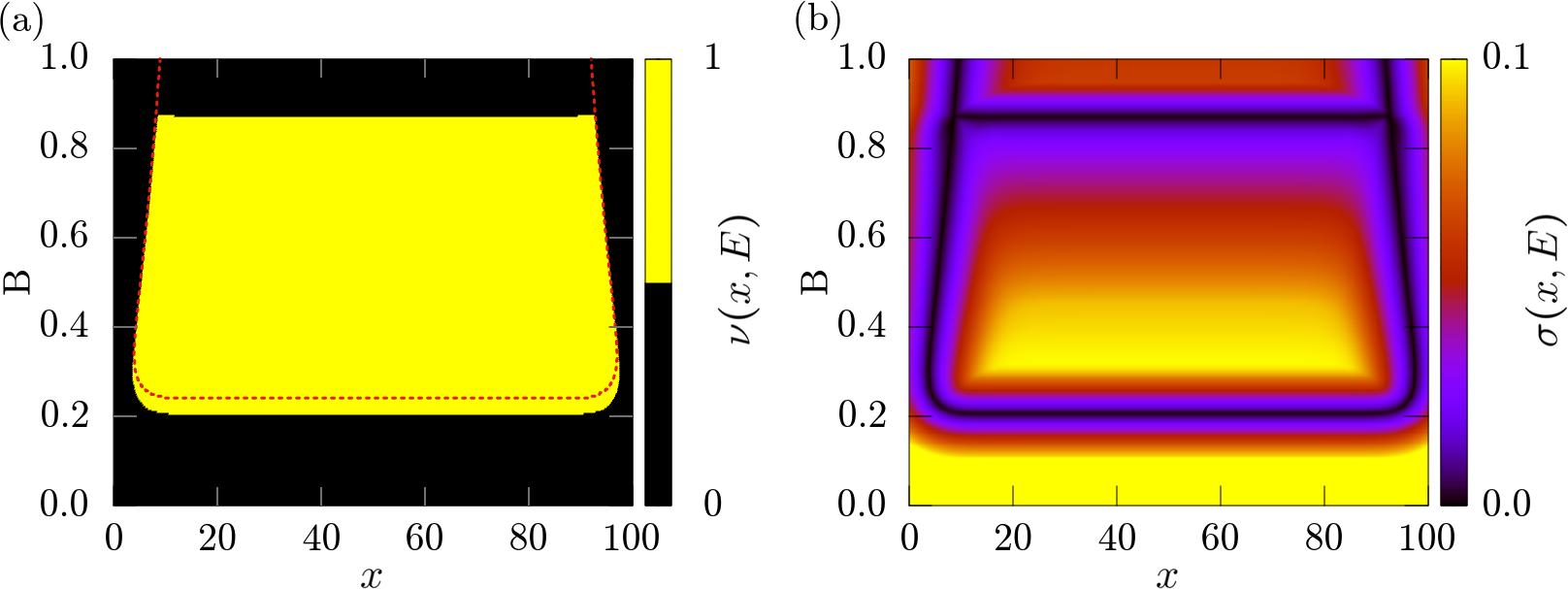}
    \caption{(a) Topological invariant $\nu(x,E)$ for $0$th photon sector ($E=\omega/2$) showing the existence of topological phase (in yellow) in the system as the spectral gap $\sigma(x,E)$ (b) is closed. 
    Red dashed line shows the outline of the topological phase for the $1$st photon sector ($E=3\omega/2$).
    Parameters are $\mu=-2t, \ \Delta=\alpha=0.2t, \ \gamma=0.2t, \ \omega=3t, \ N_{ph}=1$.}
    \label{fig:topo4}
\end{figure}

With states hybridizing between different photon sectors, it is important to ask how this influences the topology. 
In Fig.~\ref{fig:topo4}(a), we show the localizer index $\nu(x,E)$ as a function of the position $x$ and magnetic field $B$ for the 0th photon sector. 
The red dashed curve indicates the area with a non-zero topological index for the 1st photon sector. This shows that there is a mismatch between the topologically non-trivial region between the 0th and the 1st photon sector, similar to the high-frequency case.
Moreover, the inward tilt of the (almost) vertical boundaries of the topological phase for both sectors shows how the SL is sensitive to the increasing delocalization of the MBS with increasing magnetic field. This delocalization is similar in both photon sectors.
In Fig.~\ref{fig:topo4}(b), we illustrate the localizer gap $\sigma(x,E)$ as a function of position $x$ and magnetic field $B$ for the 0th photon sector. The closing of the localizer gap, i.e., $\sigma(x,E)=0$, is indicated by black color. Overall, the gap closes when the invariant is changing, as expected.

By comparing with the energy spectrum, we find that the topological transitions indicated in Figs.~\ref{fig:topo4}(a,b) are consistent with bulk gap closings and development of zero-energy states, except for the transition around $B\simeq 0.85t$ along the whole nanowire in the $0$th photon sector. Even though both SL invariant and gap here indicate a topological phase transition, the energy spectrum does not show any topological gap closings at this magnetic field but instead shows a clear permanence of the MBS within a topological gap, see Fig.~\ref{fig:topo3}(a), lower panel at $B\simeq 0.85t$. 
We attribute this problem to hybridization between states of different photon sectors. In particular, bulk states from one photon sector cross the MBS of another sector, and while there is a small hybridization in energy and states, there is clearly no bulk gap closing for the MBS to disappear and the topology to change. Still, the SL invariant and gap indicate such a change.
This leads to an erroneous \textit{pollution} of the phase diagram. We find similar pollution in all photon sectors below the highest possible sector i.e.~ for $N^{ps}<N_{ph})$.

We attribute the origin of the topological phase diagram pollution to two related elements: the low-frequency regime and the breaking of the chiral symmetry $S$ by the photon energy $\hbar \omega$ terms. 
The main problem in the low-frequency regime for the SL is the overlap of energy levels between different photon sectors, resulting in states from the continuum hybridizing with MBS energy levels. This is erroneously registered by the SL as a topological transition.
The fundamental reason for this error is that the photon energy $\hbar \omega$ terms break chiral symmetry. This chiral symmetry breaking technically invalidates the usage of the symmetry dependent approach of the SL~\cite{loring.15}. 
However, the SL formulation removes this chiral symmetry breaking term in the probed photon sector with the $-EI$ term, setting the effective photon energy in that sector to $0$. 
As a result, the SL still performs well in the high-frequency regime. 
However, in the low-frequency regime, different photon sectors overlap such that the effective cancellation of chiral symmetry breaking by the $-EI$ term is no longer present for all photon sectors present at energy $E$. This leads to an erroneous pollution of the topological phase diagram. We note that the problems with pollution is even more severe for a SL not utilizing the chiral symmetry, see Appendix \ref{app:sl}.

One simple way mitigate this effect and to probe the topology of all photon sectors safely is to engineer the spectral localizer $L_{x,E}(X,\mathcal{H}_{\infty})$ by recasting the element related to energy resolution to a new form:
\begin{eqnarray}
    \mathcal{H}_{\infty}-E I \rightarrow \mathcal{H}_{\infty}-E \, \epsilon(N^{ps}_*),
    \label{eq:fixed}
\end{eqnarray}
which changes the diagonal entries from $1$ in the identity matrix $I$ to $\epsilon(N^{ps}_*)=2\delta_{N^{ps},N^{ps}_*}-1$, with $N^{ps}_*$ being the probed photon sector. 
This has no influence on the probed photon sector, as it keeps its energy to $E$, but results in all other sectors that could cause pollution to be effectively pushed away in energy. This recasting  of the SL's energy term allows for the recovery of the unpolluted topological diagrams without any qualitative changes to the overall shape of the topological phase. In Appendix~\ref{app:sleng}, we provide details on this SL engineering and illustrate its applicability. Note that this does not shift the energy levels or remove hybridization in the energy spectrum of the Hamiltonian $\mathcal{H_1}$, but only engineers the topological probe, the SL.
This energy adjustment $\epsilon$ introduces an additional degree of freedom to the SL apart from $\kappa$, yielding the following form:
\begin{align}
L_{x,E}(X,\mathcal{H}_{\infty}) = \kappa \left(X-xI \right) \tau_x \!+\! \left(\mathcal{H}_{\infty} \! - \! E \epsilon(N^{ps}_*) \right) \tau_y.
    \label{Eq:SpectralLocalizer2}
\end{align}

\begin{figure}
    \centering
    \includegraphics[width=\linewidth]{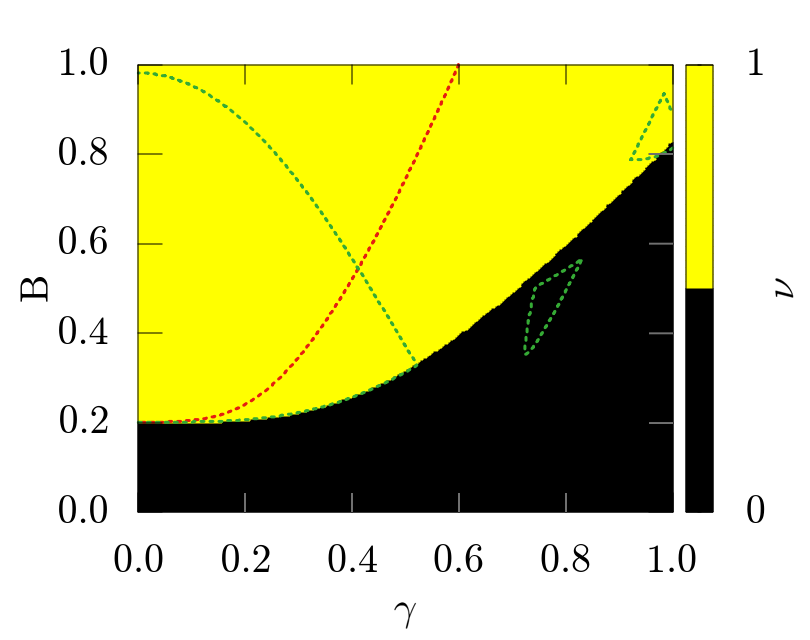}
    \caption{Topological phase diagram for $\gamma$ vs.~$B$ for the $0$th photon sector ($E=\omega/2$) using the engineered SL invariant [Eq.~\eqref{Eq:SpectralLocalizer2}]. 
    Red dashed line marks the topological phase transition in the $1$st photon sector ($E=3\omega/2$) and green lines outline the polluted topological phases stemming from photon sectors overlap in the case of non-engineered SL (cf.~with Fig.~\ref{fig:badeigen}).
    Parameters are $\mu=-2t, \ \Delta=\alpha=0.2t,  \ \omega=3t, \ N_{ph}=1$.
    }
    \label{fig:works}
\end{figure}

We show in Fig.~\ref{fig:works} the phase diagram with pollution removed by using the modified SL described by Eq.~\eqref{Eq:SpectralLocalizer2}. This diagram can be directly compared to the high-frequency regime results in Fig.~\ref{fig:topo2}(b).
As previously, colors denote the topological phase in the 0th photon sector, while the red dashed line marks the topological phase transition for the 1st photon sector. The phase diagram shows nontrivial phases analogously to the high-frequency regime, thus proving the existence of MBS for all frequencies and independent of significant hybridization between different photon sectors.
Moreover, with green dashed lines we show the polluted parts of the topological diagram, where the non-engineered SL strays away from the typical topological phases for MBS systems in 1D with a $mod(2)$ valued invariant (see Fig.~\ref{fig:badeigen}~in Appendix~\ref{app:sl} for more details). 
Overall, these results illustrate that by modifying the energy term in the SL, we shift the unprobed photon sectors in energy and thereby recover the correct information about the topological phase of the probed sector. Thus, Eq.~\eqref{Eq:SpectralLocalizer2} presents a feasible and simple workaround to extend the applicability of the SL to the realm of topology in photonic cavities at all frequencies, even in the case of overlapping photon sectors.

\section{Summary and Outlook}
\label{sec:summary}

In this work, we investigate a 1D topological superconductor (1DTSC) with Majorana bound states (MBS) inside a photonic cavity containing a finite number of photons. The resulting finite light-matter interaction has several effects on the topology and MBS of the 1DTSC: it shifts the critical magnetic field required for the topological transition in the system, and importantly, it induces parameter dependencies for the MBS energy, that we identify as a pseudo-dispersion (PD), all dependent on the photon number.
As such, the tunability is greatly enhanced, as the MBS energy can be tuned with photon number, light-matter coupling, and magnetic field. In addition, while the cavity reduces size of  the  topological gap, we find that disorder does not reduce the stability of the topological phase inside the cavity. 

As a second major result of this work, we develop an engineered spectral localizer (SL) as a versatile tool to probe topology in cavity systems. Notably, with a finite number of photons present, topology has to be probed at different energies $\hbar \omega$, with $\omega$ being the photon frequency. The SL has been developed as an outstanding tool to probe topology both in real space and with full energy resolution \cite{loring.15,CerjanJMP2023}. We find that in the high-frequency photon regime, an SL employing chiral symmetry precisely identifies all topological transitions, while also directly probing effects of disorder.
However, at low photon frequencies, the SL encounters limitations: hybridization between MBS of one photon sector with bulk states of different photon sectors leads to a erroneous pollution of the topological invariant, obscuring a clear identification of nontrivial phases. Despite this breakdown of the topological invariant, the MBSs themselves survive the hybridization with these bulk states, indicating that their topological protection persists under such conditions. To resolve this issue, we develop an engineered SL that effectively expels the protruding bulk states from the probed photon sector. This allows for efficient calculations of the correct topological diagrams, while preserving the physical behavior of the system.

Overall, our results reveal how coupling a 1DTSC to a cavity introduces new mechanisms for manipulating and stabilizing MBS, highlighting the prospective opportunities for potential experimental application.
At the same time, our results establish a versatile and computationally efficient tool to address topology in photonic cavities. We further note that recent results have shown that the superconducting gap can be enhanced in a cavity \cite{KozinPRB2025}, which opens up the possibility of further enhancing the robustness of MBS. We leave such self-consistent calculations of the superconducting order parameter to future work. 
Considering light-matter couplings in the range $\frac{\gamma}{\omega} \sim 0.1-1$ depending on materials ~\cite{anappara09,scalari12,maissen14,Gambino2014,Chikkaraddy2016,Yoshihara2017,flick17,Bayer2017,genco17,FriskKockum2019,Halbhuber2020}, our results are clearly within experimental reach. 
However, interesting questions still remain about the experimental detection of the MBS in a cavity.
There have been a few theoretical works discussing the electronic transport signature of topological states in a cavity~\cite{DavidPRL2017,NguyenPRB2024}. Thus, it would be interesting to find a transport-based signature for the MBS, capable of extracting an energy-resolved conductance to identify the MBS at different energies.

\subsection*{Acknowledgments}
% \begin{center}
%     {\small \textbf{Acknowledgments}}
% \end{center}
We thank A.~Cerjan, D.~A.~D.~Chaves, N.~C.~Costa, A.~Grushin, and D.~Svintsov for insightful discussions. 
We acknowledge financial support from the Swedish Research Council (Vetenskapsrådet) Grant No.~2022-03963 and the Knut and Alice Wallenberg Foundation, Grant Nos.~2019-0068 and 2023-0244. 
Numerical calculations were enabled by resources provided by the National Academic Infrastructure for Supercomputing in Sweden (NAISS), partially funded by the Swedish Research Council through grant agreement No.~2022-06725.

\appendix
\begin{onecolumngrid}

\section{Light-matter coupling}\label{app:matter_light} 
In this Appendix, we provide the details for the derivation of Eq.~\eqref{eq:G}, which describes the effect of light-matter coupling on the spin-conserving $t$ and spin-flipping $\alpha$ hopping amplitudes.
Since we have a system composed of fermions and bosons, it is convenient to write the Hamiltonian in a photon basis as
\begin{equation}
    {\cal H}_\infty=\sum\limits_{N}\ket{N}\bra{N} {\cal H}_C \sum\limits_{M}\ket{M}\bra{M}=\sum\limits_{N,M} \underbrace{\braket{N|{\cal H}_C|M}}_{\equiv {\cal H}_{N,M}} \ket{N}\bra{M},
\end{equation}
where $\ket{N}$ represents the photon state with $N$ photons. Note that ${\cal H}_{N,M}$, Eq.~\eqref{Eq:RashbaHamwithlight} in the main text, is still an operator in the fermionic Fock space. With this form of the Hamiltonian, we can assume that the cavity will have mainly states in a few photon sectors due to the thermal distribution and introduce a cut-off in $N=M$. Also, due to the coupling between light and matter, we have finite transition amplitudes between different photon states, i.e.~${\cal H}_{N,M}\neq \delta_{N,M}$. Below we review how to explicitly compute these matrix elements following Refs.~\cite{svintsov.alymov.24,macedofaundez24}) and apply it directly to the 1DTSC Hamiltonian~[Eq.~\eqref{eq:NormalH}] in the main text. 

We couple fermions to a photon using Peierls' substitution as shown in the main text. Since the photon creation and annihilation operators do not commute, we can re-express the dressing of the hopping terms (kinetic and SOC) using the Baker-Campbell-Hausdorff formula as
\begin{equation} 
   \exp\left(\pm \frac{i \gamma}{t} (b^{\dagger}+b)\right)= \exp\left(-\frac{\gamma^2}{2t^2}\right) \exp\left(\pm \frac{i \gamma}{t} b^{\dagger}\right)\exp\left(\pm \frac{i \gamma}{t} b\right),
\end{equation}
where we use $\left[b, b^\dagger\right]=1$. Notice that this expression is exact, since terms with multiple commutators are identically zero. 
Next, expanding the exponentials on the right hand side in terms of creation and annihilation operators,
\begin{equation} 
    \exp\left(\pm \frac{i \gamma}{t} (b^{\dagger}+b)\right)= \exp\left(-\frac{\gamma^2}{2t^2}\right) \sum\limits_{n,m=0}^{\infty}\frac{\left(\pm \frac{i \gamma}{t} \right)^{n+m}}{n!m!}b^{\dagger n}b^{m},
\end{equation}
we find that the inner product between the electron-photon coupling and the photon states is 
\begin{equation} 
    \braket{N| \exp\left(\pm \frac{i \gamma}{t} (b^{\dagger}+b)\right)|M}= \exp\left(-\frac{\gamma^2}{2t^2}\right) \sum\limits_{n,m=0}^{\infty}\frac{\left(\pm \frac{i \gamma}{t} \right)^{n+m}}{n!m!}\braket{N|b^{n}b^{\dagger m}|M}.
\end{equation}
As $b\ket{N}=\sqrt{N}\ket{N-1}$, $\bra{M}b^\dagger=\sqrt{M}\bra{M-1}$, and $\braket{N|M}=\delta_{N,M}$ 
, the inner product is given by 
\begin{eqnarray}
    \braket{N|\exp\left(\pm \frac{i \gamma}{t} (b^{\dagger}+b)\right)|M} = \exp\left(-\frac{\gamma^2}{2t^2}\right) \sum\limits_{n=0}^{N}\sum\limits_{m=0}^{M}\frac{\left(\pm \frac{i \gamma}{t} \right)^{n+m}}{n!m!}\delta_{N-n, M-m} P(n, N) P(m, M),
\end{eqnarray} 
where $P(n,N)$ is described by Eq.~\eqref{eq:poch}.
For simplicity, we define the function
\begin{equation}
    G_{N, M}(x)=\sum\limits_{n=0}^{N} \sum\limits_{m=0}^{M}\frac{\left(-i x\right)^{n+m}}{n!m!}\delta_{N-n, M-m}P(n, N)P(m, M) e^{-\frac{x^2}{2}},
\end{equation}
such that
\begin{equation} 
    \braket{N|  \exp\left(\pm \frac{i \gamma}{t} (b^{\dagger}+b)\right)|M}=G_{N,M}(\mp \gamma/t).
\end{equation}
Since aside from this electron-photon interaction, we only have the energy of the photons, $\hbar \omega \left(N+\frac{1}{2}\right) \delta_{N,M}$, to fully capture the cavity effect on the 1DTSC, we can write the Hamiltonian for the 1DTSC coupled to a photonic cavity as Eq.~\eqref{Eq:RashbaHamwithlight} in the main text.
We also note that for clarity, we omit $t$ in the related equations in the main text as it is set to be the unit of energy.

\section{Choice of spectral localizer}
\label{app:kappa}
\begin{figure}[htb]
    \centering
    \includegraphics[width=0.45\linewidth]{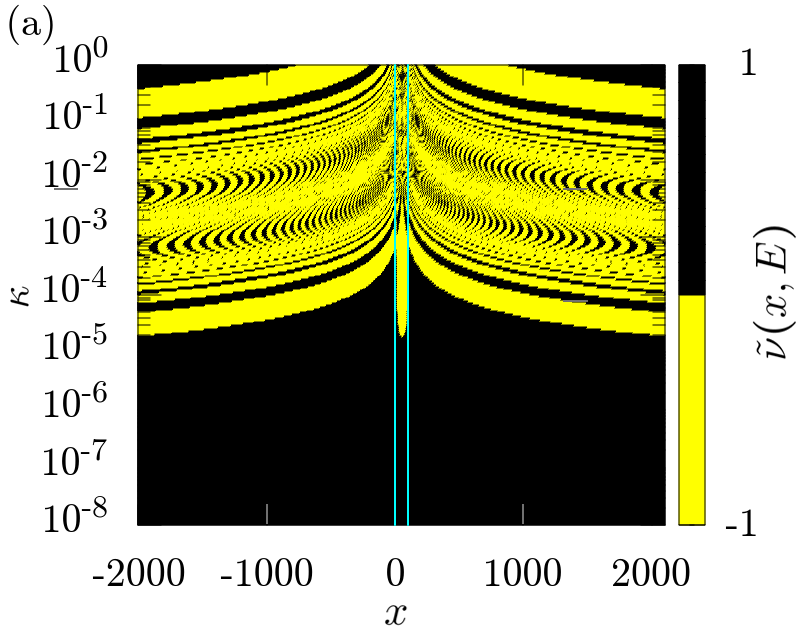}
    \includegraphics[width=0.45\linewidth]{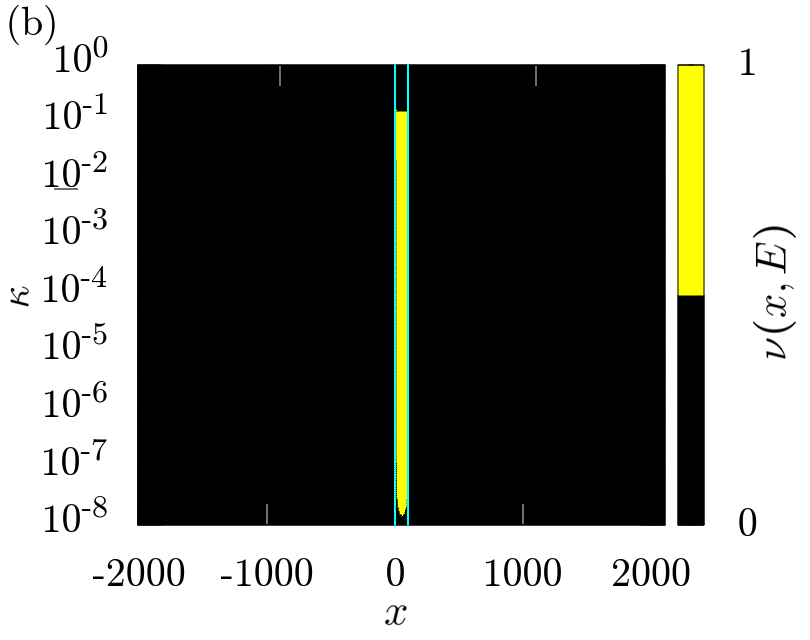}
    \caption{Topological phase diagrams in the $0$th photon sector ($E=\omega/2$) using the SL without chiral symmetry, $\nu$ (a) and with chiral symmetry as in the main text (b) (see Appendix \ref{app:sl}) along the nanowire (marked by cyan lines) with an extended vacuum region outside and as a function of $\kappa$.
    Parameters are $\mu=-2t, \ \Delta=\alpha=\gamma=0.2t, \ B=0.4t \ \omega=10t$ $N_{ph}=1$.
    }
    \label{fig:kappa}
\end{figure}

The spectral localizer (SL) is an outstanding tool to calculate topology in real space. 
It can identify the topology of a system not only within the system itself but also outside, in the vacuum.
However, understanding of topology via the SL invariant might be obfuscated by anomalous results stemming from broken symmetries.
In particular, we analyze two topological diagrams produced by separate versions of the SL invariant:
\begin{align}
    {\rm (a)}&:  \tilde{\nu}(x,E)={\rm sign} \det[\tilde{X}+i (\mathcal{H}_{1}-EI)],~{\rm Figs.~\ref{fig:kappa}(a), Fig.~\ref{fig:badSL}},  \\
     {\rm (b)}&:  \nu(x,E)={\rm sig}([\tilde{X}+i (\mathcal{H}_{1}-EI)]S),~{\rm main \, text, \, Fig.~\ref{fig:kappa}(b), \, Fig.~\ref{fig:badeigen}}.
\end{align}
We use $\nu(x,E)$ as a topological indicator throughout the main text, thus exploiting the chiral symmetry of the system.

In this Appendix, we analyze how the SL topological invariant depends on the parameter $\kappa$ [Eq.~\eqref{Eq:SpectralLocalizer}]. This supports the choice of $\kappa = 10^{-2}$ in the main text, and also illustrates how the topological invariant $\nu$, exploiting chiral symmetry used in the main text, is superior to $\tilde{\nu}$, extracted without chiral symmetry.
In Fig.~\ref{fig:kappa} we plot $\tilde{\nu}$ (a) and $\nu$ (b) in the nanowire, but also in an extended (vacuum) area around it (cyan lines mark the nanowire) for varying $\kappa$.
The results in Fig.~\ref{fig:kappa} shows the clear advantage of employing the chiral symmetry $S$ version of the SL: for $\nu$ we correctly capture the topological phase inside the nanowire and a trivial phase outside for a range of $\kappa$ spanning seven orders of magnitude, including the value used in the main text. In contrast, for the SL without chiral symmetry enforced, $\tilde{\nu}$, SL incorrectly predicts a non-trivial vacuum for $\kappa \gtrsim 10^{-5}$ in the region of $\pm 2000$ sites near the nanowire.
Note here that as the invariant $\tilde{\nu}$ is calculated with a formula lacking the chiral operator $S$, the topologically nontrivial phase is denoted by $-1$ (yellow) and the trivial one with $1$ (black), in contrast values $0,1$ of the topological indicator in the rest of the results in this work. In order to maintain visual coherence, colors denote the same topological phases throughout the entire manuscript.

\section{Pollution of the topological phase diagram}
\label{app:sl}
\begin{figure}
    \centering
    \includegraphics[width=\linewidth]{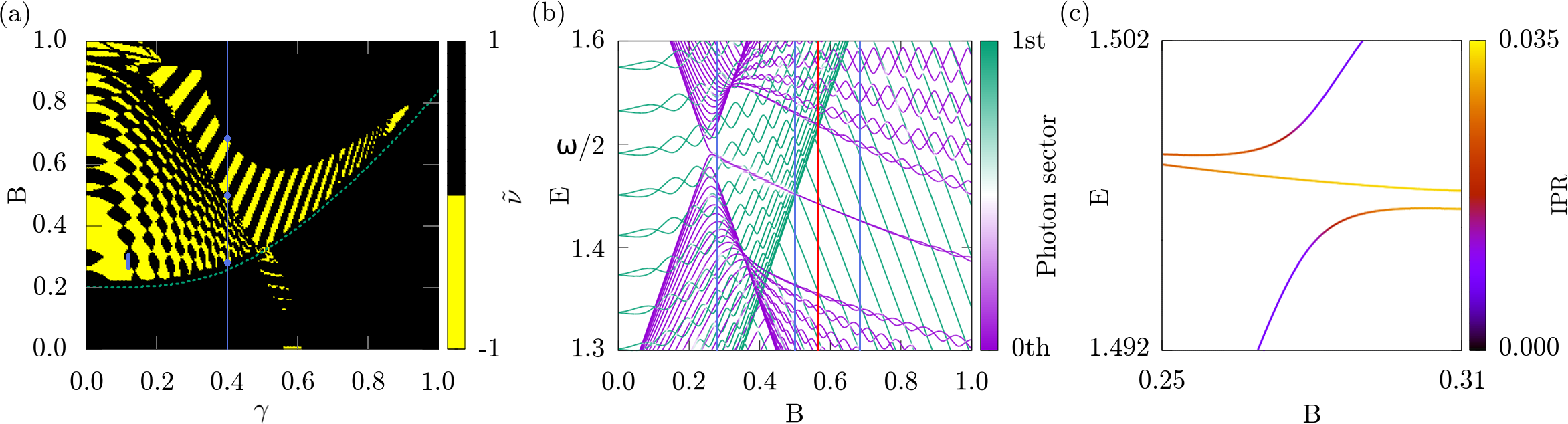}
    \caption{
    Topological diagram of $0$th photon sector using the $\tilde{\nu}$ topological invariant (a) with corresponding energy spectrum (b) along the blue line in (a). Blue dots mark apparent topological transitions and are also marked by blue lines in (b).
     Green dashed line shows the boundary of the topological phase in the pollution-free high-frequency regime ($\omega = 5t$). 
     (c) Zoomed-in energy spectrum displaying single hybridization of MBS with a bulk state at the blue rectangle in (a), with colors showing the IPR. Red line in (b) corresponds to a red marks in Fig.~\ref{fig:badeigen} near $B \simeq 0.57t$.
    Parameters are $\mu=-2t, \ \Delta=\alpha=0.2t, \ \omega=3t, \ N_{ph}=1$, and $\kappa=10^{-4}$ to maximize the stability, and (b) $ \gamma=0.4t, $ (c) $ \gamma=0.12t$.
    }
    \label{fig:badSL}
\end{figure}

In this Appendix, we provide an expansion on the discussions in the main text on both the choice of SL invariant and the issue of pollution of the phase diagram. 
When the separation between photon sectors is large enough to prevent bulk states of one photon sector from entering the topological gap of another, the SL always correctly predicts the topological phases.
However, understanding of topology via the SL invariant is hindered by the anomalous results it produces in the low-frequency regime, resulting in \textit{pollution} of the topological diagram by artifacts showing nonphysical values, i.e.~beyond the $0, 1$ values of the topological invariant $\nu$. 
We start by investigating the SL without exploiting chiral symmetry. Note that as this invariant, $\tilde{\nu}$, is calculated with a formula lacking the chiral operator $S$, the topologically nontrivial phase is denoted by $-1$ (yellow) and the trivial one with $1$ (black), in contrast to the rest of the results in this work. 
In Fig.~\ref{fig:badSL}(a), we plot the topological indicator $\tilde{\nu}$ as a function of $\gamma$ and $B$, which displays a complicated pattern of checkerboard and striped regions of interwoven topological phases. To understand these patterns, we plot the energy spectrum in Fig.~\ref{fig:badSL}(b) for $\gamma =0.4$ along the blue line in (a), with the dots corresponding blue lines in (b).
We also mark with a green dashed line the pollution-free topological phase transition in the high-frequency regime as a guide to the eye.

We find that the initial transition with increasing $B$ into the topological regime is marked by the development of the black-yellow interwoven pattern. The pattern is due to the topological invariant picking up each crossing of the MBS in the 0th photon sector (violet) with bulk states from the 1st sector (green). 
The checkerboard pattern is due to the bulk states displaying a typical oscillatory pattern as a function of $B$, while the stripe pattern appears once the bulk states have a linear dependence on $B$ when they cross the MBS.
In Fig.~\ref{fig:badSL}(c), we zoom in on one of these crossings [blue rectangle in (a)]. As seen, there is a clear energy hybridization between the near constant-energy MBS and the dispersive bulk states. By utilizing the inverse participation ratio (IPR)~\cite{thouless.74} (colors), we also see that the near constant-energy states stays localized, as expected for MBS.
As a result, the MBS spectral weight is distributed across three states rather than the original two MBS (one at each end of the nanowire). This hybridization alters the topological invariant, producing checkerboard or striped features in the topological diagram. We find similar checkerboards/striped patterns when bulk states from one photon sector overlap with the MBS of another sector for all investigated sectors and frequencies ($\omega \lesssim 5t$).

Turning to the case of $\nu$, the topological indicator constrained by the chiral $S$ operator, we show its topological phase diagram in Fig.~\ref{fig:badeigen}(a). 
Here, even if the interwoven patterns disappear, we still find several topological phases emerging, while the only physically  meaningful topological gap closing should occur along the green dashed line of the pollution-free high-frequency regime. 
Notably, we also expect the topological invariant to only take values $0,1$ (black, yellow).  This shows that the chiral-symmetry enforced SL is superior, but also that photon sector pollution is still an issue. The pollution occurs in all but the highest photon sector (only the 0th sector is shown in (a), but for any $N_{ph}$, only the $N_{ph}$th photon sector is free from pollution). 
We aim to further understand this pollution by extracting the eigenvalues of the SL in Fig.~\ref{fig:badeigen}(b) along the red line in (a) with dots marking red lines in (b). Since the invariant $\nu$ is defined as the signature of the SL matrix, the topological invariant ${\nu}$ reflects the imbalance between SL eigenvalues above and below zero. Thus, the zero-crossings at the two red lines in (b) change ${\nu}$. The red line at lower $B$ correctly indicates the topological gap closing in the energy spectrum. However, for the red line at the higher $B \sim 0.57$, we cannot extract any specific feature in the energy spectrum, which corresponds to the red line in the energy spectrum in Fig.~\ref{fig:badSL}(b). 
Note here that the crossing of the edge of the gap of $0$th photon sector with the bulk states of $1$st is accidentally occuring at the red line in Fig.~\ref{fig:badSL}(b) and will not appear for another $\gamma$. It is therefore unclear why the SL eigenspectrum crosses zero at this magnetic field value. 
However, in the main text, we are able to correct for these issues and extract the correct phase diagram in Fig.~\ref{fig:works} by engineering the SL such that it effectively shifts irrelevant photon sectors in energy.

\begin{figure}
    \centering
    \includegraphics[width=0.85\linewidth]{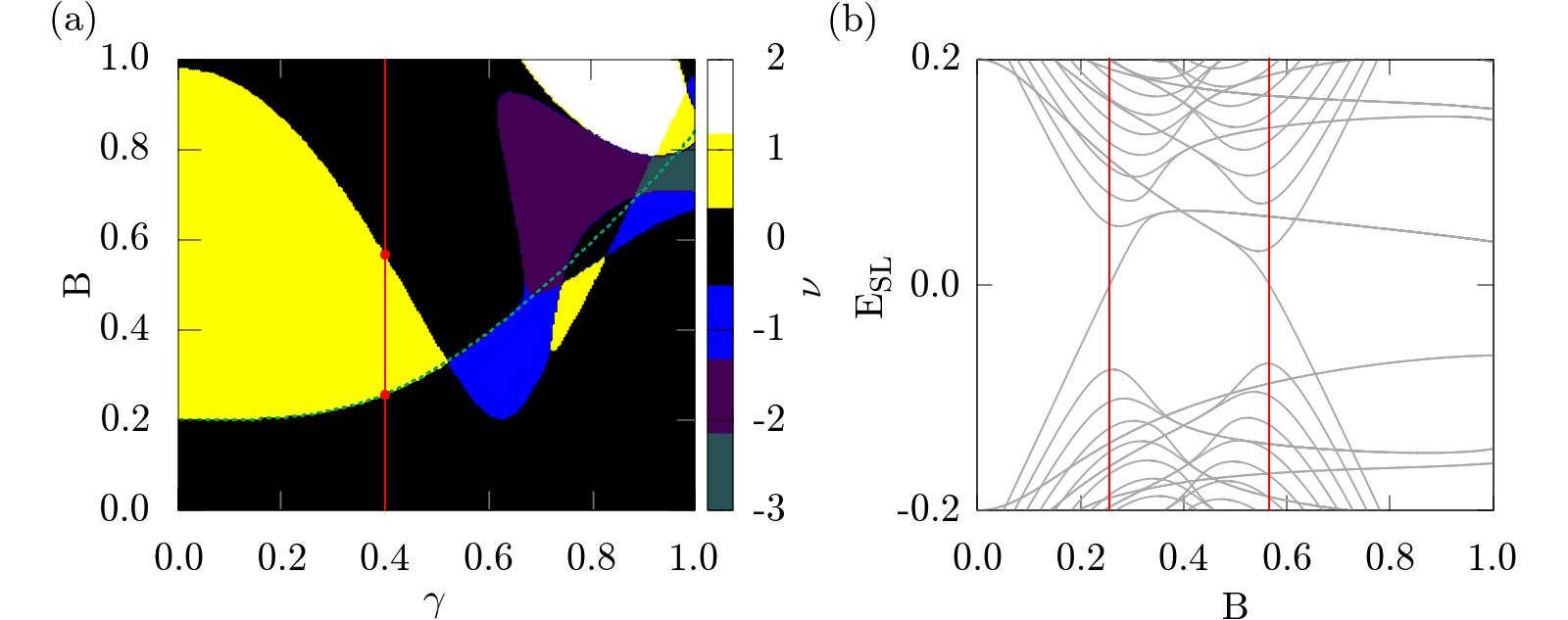}
    \caption{
    Topological diagram of $0$th photon sector using the $\nu$ topological invariant with chiral symmetry (a) with corresponding spectral localizer eigenspectrum $E_{SL}$ (b) along the red line in (a). 
    Red dots mark apparent topological transitions and are also marked by red lines in (b).
    Green dashed line shows the boundary of the topological phase in the pollution-free high-frequency regime ($\omega = 5t$). 
    Parameters are $\mu=-2t, \ \Delta=\alpha=0.2t, \ \omega=3t, \ N_{ph}=1, \  \kappa=10^{-2}$ (b) $ \gamma=0.4t$. }
    \label{fig:badeigen}
\end{figure}

\section{Engineering the spectral localizer}
\label{app:sleng}
In order to fix the \textbf{(a)} (a) pollution of the topological diagram in the low-frequency regime and thus to probe the topology of all photon sectors without issues, we engineer the spectral localizer $L_{x,E}(X,\mathcal{H}_{\infty})$ by modifying the energy term in Eq.~\eqref{eq:fixed} in the main text. In this Appendix, we provide additional details. 
Equation~\eqref{eq:fixed} amounts to, when probing topology in the $N^{ps}_*$ photon sector, using
\begin{eqnarray}
    \mathcal{H}_{\infty} - E I  \rightarrow 
    \begin{pmatrix}
         {\cal H}_{0,0} & {\cal H}_{0,1} & {\cal H}_{0,2} & \cdots  \\
         {\cal H}_{1,0} & {\cal H}_{1,1} & {\cal H}_{1,2} & \cdots  \\ 
         {\cal H}_{2,0} & {\cal H}_{2,1} & {\cal H}_{2,2} & \cdots  \\
         \vdots  & \vdots  & \vdots  &  \ddots  
    \end{pmatrix}& 
    - E
    \begin{pmatrix}
         \epsilon(N^{ps}_*) & 0 & 0 & \cdots  \\
         0 & \epsilon(N^{ps}_*) & 0 & \cdots  \\ 
         0 & 0 & \epsilon(N^{ps}_*) & \cdots  \\
         \vdots  & \vdots  & \vdots  &  \ddots  
    \end{pmatrix}, 
    \label{eq:fixed2}
\end{eqnarray}
with  $\epsilon(N^{ps}_*) = 2 \delta_{N^{ps},N_*^{ps}}-1$. This keeps the diagonal to $-E$ for the probed photon sector as in the original SL, but flips the signs for the energy in all other sectors, whose bulk spectrum may pollute the topological invariant. This way we distance the probed sector from all other sectors by at least $2\omega$ in the SL, which we find removes all issues with the topological invariant, see Fig.~\ref{fig:works} in the main text. 

We note that this modification is linked to the frequency $\omega$, as that sets the energy spacing between the centers of each photon sector. 
In order to clarify our choice of $\epsilon(N^{ps})=-1$ for the unprobed photon sectors, we show in Fig.~\ref{fig:appeps} a heatmap of the minimum (a) and maximum (b) values of the topological indicator $\nu$ in the entire nanowire for different photon energies $\omega$ and different energy modification $\epsilon(N^{ps})$. 
The parameters are chosen to focus on one of the many problematic points in the polluted topological diagram where $\nu=2$ ($\gamma=B=0.9t$) for the non-engineered SL i.e.~when using $\epsilon(N^{ps})=1$.
This non-engineered SL is indicated with a red line, while the implemented engineered shift, $\epsilon(N^{ps})=-1$ is indicated with a green line. 
As seen, the original SL experiences pollution for a range of frequencies $\omega$, with the minimum value going below $-1$ and the maximum going above $1$, neither of which is a valid topological invariant for our 1DTSC. 
However, any choice of $\epsilon(N^{ps})\leq 0$ remedies this fact, and therefore the choice $\epsilon(N^{ps})=-1$ is both sufficient with margin and simple.
Overall, the relationship between $\omega$ and $\epsilon(N^{ps})$ can be understood by noting that the shift in energy $\epsilon(N^{ps})$ needs to be larger for smaller frequencies $\omega$ to remove overlap between different photon sectors.

\begin{figure}
    \centering
    \includegraphics[width=0.49\linewidth]{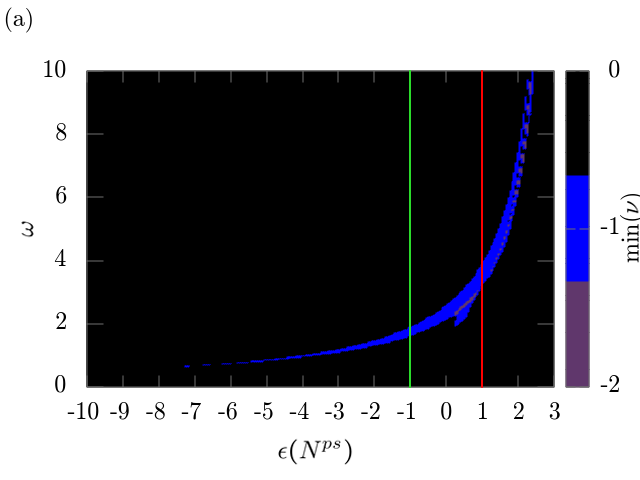}
    \includegraphics[width=0.49\linewidth]{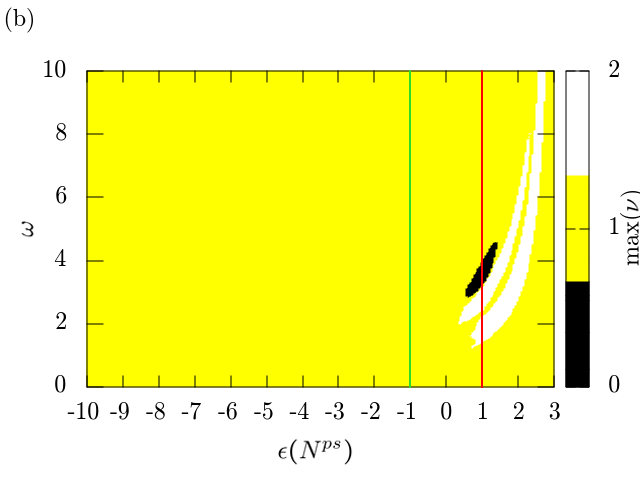}
    \caption{Heatmap of minimum (a) and maximum (b) values of the topological indicator $\nu$  for given photon energy $\omega$ and energy shift $\epsilon(N^{ps})$ of each non-probed photon sector. Colors follows the extended color scheme on Fig.~\ref{fig:badeigen}(a).
    Red line at $\epsilon(N^{ps})=1$ marks the case of non-engineered SL, green line marks the choice $\epsilon(N^{ps})=-1$ in the main text.
    Parameters are $\mu=-2t, \ \Delta=\alpha=0.2t, \ \gamma=B=0.9t, \  \kappa=10^{-2} \ N_{ph}=1$.
    } 
    \label{fig:appeps}
\end{figure}

\end{onecolumngrid}
\begin{twocolumngrid}
\end{twocolumngrid}
\bibliographystyle{apsrev4-2mod}
\bibliography{biblio}

@article{GiannettiPumpProbe2016,
    author = {Claudio Giannetti and Massimo Capone and Daniele Fausti and Michele Fabrizio and Fulvio Parmigiani and Dragan Mihailovic},
    title = {Ultrafast optical spectroscopy of strongly correlated materials and high-temperature superconductors: a non-equilibrium approach},
    journal = {Adv. Phys.},
    volume = {65},
    number = {2},
    pages = {58--238},
    year = {2016},
    publisher = {Taylor \& Francis},
    doi = {10.1080/00018732.2016.1194044},
    URL = {https://doi.org/10.1080/00018732.2016.1194044}
}

@Article{BaoNRP2022,
    author={Bao, Changhua
    and Tang, Peizhe
    and Sun, Dong
    and Zhou, Shuyun},
    title={Light-induced emergent phenomena in 2D materials and topological materials},
    journal={Nat. Rev. Phys.},
    year={2022},
    month={Jan},
    day={01},
    volume={4},
    number={1},
    pages={33-48},
    issn={2522-5820},
    doi={10.1038/s42254-021-00388-1},
    url={https://doi.org/10.1038/s42254-021-00388-1}
}

@Article{BasovNM2017,
    author={Basov, D. N.
    and Averitt, R. D.
    and Hsieh, D.},
    title={Towards properties on demand in quantum materials},
    journal={Nat. Mater.},
    year={2017},
    month={Nov},
    day={01},
    volume={16},
    number={11},
    pages={1077-1088},
    issn={1476-4660},
    doi={10.1038/nmat5017},
    url={https://doi.org/10.1038/nmat5017}
}

@article{RaimondRMP2001,
  title = {Manipulating quantum entanglement with atoms and photons in a cavity},
  author = {Raimond, J. M. and Brune, M. and Haroche, S.},
  journal = {Rev. Mod. Phys.},
  volume = {73},
  issue = {3},
  pages = {565--582},
  numpages = {0},
  year = {2001},
  month = {Aug},
  publisher = {American Physical Society},
  doi = {10.1103/RevModPhys.73.565},
  url = {https://link.aps.org/doi/10.1103/RevModPhys.73.565}
}

@article{LeibfriedRMP2003,
  title = {Quantum dynamics of single trapped ions},
  author = {Leibfried, D. and Blatt, R. and Monroe, C. and Wineland, D.},
  journal = {Rev. Mod. Phys.},
  volume = {75},
  issue = {1},
  pages = {281--324},
  numpages = {0},
  year = {2003},
  month = {Mar},
  publisher = {American Physical Society},
  doi = {10.1103/RevModPhys.75.281},
  url = {https://link.aps.org/doi/10.1103/RevModPhys.75.281}
}

@article{ManfredScience1998,
    author = {Manfred Fiebig  and Kenjiro Miyano  and Yasuhide Tomioka  and Yoshinori Tokura },
    title = {Visualization of the Local Insulator-Metal Transition in  {$\mathrm{Pr_{0.7}Ca_{0.3}MnO_3}$}},
    journal = {Science},
    volume = {280},
    number = {5371},
    pages = {1925-1928},
    year = {1998},
    doi = {10.1126/science.280.5371.1925},
    URL = {https://www.science.org/doi/abs/10.1126/science.280.5371.1925}
}

@Article{XuNL2010,
    author={Xu, Xiaodong
    and Gabor, Nathaniel M.
    and Alden, Jonathan S.
    and van der Zande, Arend M.
    and McEuen, Paul L.},
    title={Photo-Thermoelectric Effect at a Graphene Interface Junction},
    journal={Nano Lett.},
    year={2010},
    month={Feb},
    day={10},
    publisher={American Chemical Society},
    volume={10},
    number={2},
    pages={562-566},
    issn={1530-6984},
    doi={10.1021/nl903451y},
    url={https://doi.org/10.1021/nl903451y}
}

@article{OkaPRB2009,
  title = {Photovoltaic Hall effect in graphene},
  author = {Oka, Takashi and Aoki, Hideo},
  journal = {Phys. Rev. B},
  volume = {79},
  issue = {8},
  pages = {081406},
  numpages = {4},
  year = {2009},
  month = {Feb},
  publisher = {American Physical Society},
  doi = {10.1103/PhysRevB.79.081406},
  url = {https://link.aps.org/doi/10.1103/PhysRevB.79.081406}
}

@article{UsajPRB2014,
  title = {Irradiated graphene as a tunable Floquet topological insulator},
  author = {Usaj, Gonzalo and Perez-Piskunow, P. M. and Foa Torres, L. E. F. and Balseiro, C. A.},
  journal = {Phys. Rev. B},
  volume = {90},
  issue = {11},
  pages = {115423},
  numpages = {12},
  year = {2014},
  month = {Sep},
  publisher = {American Physical Society},
  doi = {10.1103/PhysRevB.90.115423},
  url = {https://link.aps.org/doi/10.1103/PhysRevB.90.115423}
}

@article{FloquetGuPRL2011,
  title = {Floquet Spectrum and Transport through an Irradiated Graphene Ribbon},
  author = {Gu, Zhenghao and Fertig, H. A. and Arovas, Daniel P. and Auerbach, Assa},
  journal = {Phys. Rev. Lett.},
  volume = {107},
  issue = {21},
  pages = {216601},
  numpages = {5},
  year = {2011},
  month = {Nov},
  publisher = {American Physical Society},
  doi = {10.1103/PhysRevLett.107.216601},
  url = {https://link.aps.org/doi/10.1103/PhysRevLett.107.216601}
}

@article{Eckardt2017,
  title = {{Colloquium: Atomic quantum gases in periodically driven optical lattices}},
  author = {Eckardt, Andr\'e},
  journal = {Rev. Mod. Phys.},
  volume = {89},
  issue = {1},
  pages = {011004},
  numpages = {30},
  year = {2017},
  month = {Mar},
  publisher = {American Physical Society},
  doi = {10.1103/RevModPhys.89.011004},
  url = {https://link.aps.org/doi/10.1103/RevModPhys.89.011004}
}

@article{oka2019,
	author = {Oka, Takashi and Kitamura, Sota},
	title = {{Floquet Engineering of Quantum Materials}},
	journal = {Annu. Rev. Condens. Matter Phys.},
	volume = {10},
	number = {1},
	pages = {387-408},
	year = {2019},
	doi = {10.1146/annurev-conmatphys-031218-013423}
}

@article{GhoshJPCMReview2024,
    doi = {10.1088/1361-648X/ad0e2d},
    url = {https://dx.doi.org/10.1088/1361-648X/ad0e2d},
    year = {2023},
    month = {nov},
    publisher = {IOP Publishing},
    volume = {36},
    number = {9},
    pages = {093001},
    author = {Arnob Kumar Ghosh and Tanay Nag and Arijit Saha},
    title = {Generation of higher-order topological insulators using periodic driving},
    journal = {J. Phys.: Condens. Matter}
}

@article{FaustiScience2011,
    author = {D. Fausti  and R. I. Tobey  and N. Dean  and S. Kaiser  and A. Dienst  and M. C. Hoffmann  and S. Pyon  and T. Takayama  and H. Takagi  and A. Cavalleri },
    title = {Light-Induced Superconductivity in a Stripe-Ordered Cuprate},
    journal = {Science},
    volume = {331},
    number = {6014},
    pages = {189-191},
    year = {2011},
    doi = {10.1126/science.1197294},
    URL = {https://www.science.org/doi/abs/10.1126/science.1197294}
}

@article{RitschRMP2013,
  title = {Cold atoms in cavity-generated dynamical optical potentials},
  author = {Ritsch, Helmut and Domokos, Peter and Brennecke, Ferdinand and Esslinger, Tilman},
  journal = {Rev. Mod. Phys.},
  volume = {85},
  issue = {2},
  pages = {553--601},
  numpages = {0},
  year = {2013},
  month = {Apr},
  publisher = {American Physical Society},
  doi = {10.1103/RevModPhys.85.553},
  url = {https://link.aps.org/doi/10.1103/RevModPhys.85.553}
}

@article{SchlawinAPR2022,
    author = {Schlawin, F. and Kennes, D. M. and Sentef, M. A.},
    title = {Cavity quantum materials},
    journal = {Appl. Phys. Rev.},
    volume = {9},
    number = {1},
    pages = {011312},
    year = {2022},
    month = {02},
    issn = {1931-9401},
    doi = {10.1063/5.0083825},
    url = {https://doi.org/10.1063/5.0083825}
}

@Article{LeonardNature2017,
    author={L{\'e}onard, Julian
    and Morales, Andrea
    and Zupancic, Philip
    and Esslinger, Tilman
    and Donner, Tobias},
    title={Supersolid formation in a quantum gas breaking a continuous translational symmetry},
    journal={Nature},
    year={2017},
    month={Mar},
    day={01},
    volume={543},
    number={7643},
    pages={87-90},
    issn={1476-4687},
    doi={10.1038/nature21067},
    url={https://doi.org/10.1038/nature21067}
}

@article{LeonardScience2017,
    author = {Julian Léonard  and Andrea Morales  and Philip Zupancic  and Tobias Donner  and Tilman Esslinger },
    title = {Monitoring and manipulating Higgs and Goldstone modes in a supersolid quantum gas},
    journal = {Science},
    volume = {358},
    number = {6369},
    pages = {1415-1418},
    year = {2017},
    doi = {10.1126/science.aan2608},
    URL = {https://www.science.org/doi/abs/10.1126/science.aan2608}
}

@article{SchlawinPRL2019,
  title = {Cavity-Mediated Electron-Photon Superconductivity},
  author = {Schlawin, Frank and Cavalleri, Andrea and Jaksch, Dieter},
  journal = {Phys. Rev. Lett.},
  volume = {122},
  issue = {13},
  pages = {133602},
  numpages = {6},
  year = {2019},
  month = {Apr},
  publisher = {American Physical Society},
  doi = {10.1103/PhysRevLett.122.133602},
  url = {https://link.aps.org/doi/10.1103/PhysRevLett.122.133602}
}

@article{CurtisPRL2019,
  title = {Cavity Quantum Eliashberg Enhancement of Superconductivity},
  author = {Curtis, Jonathan B. and Raines, Zachary M. and Allocca, Andrew A. and Hafezi, Mohammad and Galitski, Victor M.},
  journal = {Phys. Rev. Lett.},
  volume = {122},
  issue = {16},
  pages = {167002},
  numpages = {5},
  year = {2019},
  month = {Apr},
  publisher = {American Physical Society},
  doi = {10.1103/PhysRevLett.122.167002},
  url = {https://link.aps.org/doi/10.1103/PhysRevLett.122.167002}
}

@article{DavidPRL2017,
  title = {Cavity-Enhanced Transport of Charge},
  author = {Hagenm\"uller, David and Schachenmayer, Johannes and Sch\"utz, Stefan and Genes, Claudiu and Pupillo, Guido},
  journal = {Phys. Rev. Lett.},
  volume = {119},
  issue = {22},
  pages = {223601},
  numpages = {6},
  year = {2017},
  month = {Nov},
  publisher = {American Physical Society},
  doi = {10.1103/PhysRevLett.119.223601},
  url = {https://link.aps.org/doi/10.1103/PhysRevLett.119.223601}
}

@article{svintsov.alymov.24,
  title = {One-dimensional electron localization in semiconductors coupled to electromagnetic cavities},
  author = {Svintsov, Dmitry and Alymov, Georgy and Devizorova, Zhanna and Martin-Moreno, Luis},
  journal = {Phys. Rev. B},
  volume = {109},
  issue = {4},
  pages = {045432},
  numpages = {12},
  year = {2024},
  month = {Jan},
  publisher = {American Physical Society},
  doi = {10.1103/PhysRevB.109.045432},
  url = {https://link.aps.org/doi/10.1103/PhysRevB.109.045432}
}

@article{macedofaundez24,
  title = {Multifractal critical phase driven by coupling quasiperiodic systems to electromagnetic cavities},
  author = {Macedo, Thales F. and Fa\'undez, Juli\'an and dos Santos, Raimundo R. and Costa, Natanael C. and Pinheiro, Felipe A.},
  journal = {Phys. Rev. B},
  volume = {112},
  issue = {17},
  pages = {174202},
  numpages = {12},
  year = {2025},
  month = {Nov},
  publisher = {American Physical Society},
  doi = {10.1103/gt7l-pyql},
  url = {https://link.aps.org/doi/10.1103/gt7l-pyql}
}

@article{BacciconiPRX2025,
  title = {Theory of Fractional Quantum Hall Liquids Coupled to Quantum Light and Emergent Graviton-Polaritons},
  author = {Bacciconi, Zeno and Xavier, Hernan B. and Carusotto, Iacopo and Chanda, Titas and Dalmonte, Marcello},
  journal = {Phys. Rev. X},
  volume = {15},
  issue = {2},
  pages = {021027},
  numpages = {35},
  year = {2025},
  month = {Apr},
  publisher = {American Physical Society},
  doi = {10.1103/PhysRevX.15.021027},
  url = {https://link.aps.org/doi/10.1103/PhysRevX.15.021027}
}

@Article{ChiocchettaNC2021,
    author={Chiocchetta, Alessio
    and Kiese, Dominik
    and Zelle, Carl Philipp
    and Piazza, Francesco
    and Diehl, Sebastian},
    title={Cavity-induced quantum spin liquids},
    journal={Nat. Commun.},
    year={2021},
    month={Oct},
    day={08},
    volume={12},
    number={1},
    pages={5901},
    issn={2041-1723},
    doi={10.1038/s41467-021-26076-3},
    url={https://doi.org/10.1038/s41467-021-26076-3}
}

@article{PerezGonzalez2025lightmatter,
  doi = {10.22331/q-2025-02-17-1633},
  url = {https://doi.org/10.22331/q-2025-02-17-1633},
  title = {Light-matter correlations in {Q}uantum {F}loquet engineering of cavity quantum materials},
  author = {P{\'{e}}rez-Gonz{\'{a}}lez, Beatriz and Platero, Gloria and Gomez-Le{\'{o}}n, {\'{A}}lvaro},
  journal = {{Quantum}},
  issn = {2521-327X},
  publisher = {{Verein zur F{\"{o}}rderung des Open Access Publizierens in den Quantenwissenschaften}},
  volume = {9},
  pages = {1633},
  month = feb,
  year = {2025}
}

@article{KozinPRB2025,
  title = {Cavity-enhanced superconductivity via band engineering},
  author = {Kozin, Valerii K. and Thingstad, Even and Loss, Daniel and Klinovaja, Jelena},
  journal = {Phys. Rev. B},
  volume = {111},
  issue = {3},
  pages = {035410},
  numpages = {9},
  year = {2025},
  month = {Jan},
  publisher = {American Physical Society},
  doi = {10.1103/PhysRevB.111.035410},
  url = {https://link.aps.org/doi/10.1103/PhysRevB.111.035410}
}

@article{SentefPRR2020,
  title = {Quantum to classical crossover of Floquet engineering in correlated quantum systems},
  author = {Sentef, Michael A. and Li, Jiajun and K\"unzel, Fabian and Eckstein, Martin},
  journal = {Phys. Rev. Res.},
  volume = {2},
  issue = {3},
  pages = {033033},
  numpages = {18},
  year = {2020},
  month = {Jul},
  publisher = {American Physical Society},
  doi = {10.1103/PhysRevResearch.2.033033},
  url = {https://link.aps.org/doi/10.1103/PhysRevResearch.2.033033}
}

@article{JiajunLiPRB2022,
  title = {Effective theory of lattice electrons strongly coupled to quantum electromagnetic fields},
  author = {Li, Jiajun and Schamri\ss{}, Lukas and Eckstein, Martin},
  journal = {Phys. Rev. B},
  volume = {105},
  issue = {16},
  pages = {165121},
  numpages = {16},
  year = {2022},
  month = {Apr},
  publisher = {American Physical Society},
  doi = {10.1103/PhysRevB.105.165121},
  url = {https://link.aps.org/doi/10.1103/PhysRevB.105.165121}
}

@article{JiajunLiPRL2020,
  title = {Manipulating Intertwined Orders in Solids with Quantum Light},
  author = {Li, Jiajun and Eckstein, Martin},
  journal = {Phys. Rev. Lett.},
  volume = {125},
  issue = {21},
  pages = {217402},
  numpages = {7},
  year = {2020},
  month = {Nov},
  publisher = {American Physical Society},
  doi = {10.1103/PhysRevLett.125.217402},
  url = {https://link.aps.org/doi/10.1103/PhysRevLett.125.217402}
}

@article{gomez_dmytruk_25,
  title = {Majorana bound states from cavity embedding in an interacting two-site Kitaev chain},
  author = {G\'omez-Le\'on, \'Alvaro and Schir\`o, Marco and Dmytruk, Olesia},
  journal = {Phys. Rev. B},
  volume = {111},
  issue = {15},
  pages = {155410},
  numpages = {11},
  year = {2025},
  month = {Apr},
  publisher = {American Physical Society},
  doi = {10.1103/PhysRevB.111.155410},
  url = {https://link.aps.org/doi/10.1103/PhysRevB.111.155410}
}

@article{Dmytruk.Marco.24,
  title = {Hybrid light-matter states in topological superconductors coupled to cavity photons},
  author = {Dmytruk, Olesia and Schir\`o, Marco},
  journal = {Phys. Rev. B},
  volume = {110},
  issue = {7},
  pages = {075416},
  numpages = {10},
  year = {2024},
  month = {Aug},
  publisher = {American Physical Society},
  doi = {10.1103/PhysRevB.110.075416},
  url = {https://link.aps.org/doi/10.1103/PhysRevB.110.075416}
}

@article{BacciconiPRB2024,
  title = {Topological protection of Majorana polaritons in a cavity},
  author = {Bacciconi, Zeno and Andolina, Gian Marcello and Mora, Christophe},
  journal = {Phys. Rev. B},
  volume = {109},
  issue = {16},
  pages = {165434},
  numpages = {7},
  year = {2024},
  month = {Apr},
  publisher = {American Physical Society},
  doi = {10.1103/PhysRevB.109.165434},
  url = {https://link.aps.org/doi/10.1103/PhysRevB.109.165434}
}

@article{DartiailhPRL2017,
  title = {Direct Cavity Detection of Majorana Pairs},
  author = {Dartiailh, Matthieu C. and Kontos, Takis and Dou\ifmmode \mbox{\c{c}}\else \c{c}\fi{}ot, Benoit and Cottet, Audrey},
  journal = {Phys. Rev. Lett.},
  volume = {118},
  issue = {12},
  pages = {126803},
  numpages = {7},
  year = {2017},
  month = {Mar},
  publisher = {American Physical Society},
  doi = {10.1103/PhysRevLett.118.126803},
  url = {https://link.aps.org/doi/10.1103/PhysRevLett.118.126803}
}

@article{TrifPRL2012,
  title = {Resonantly Tunable Majorana Polariton in a Microwave Cavity},
  author = {Trif, Mircea and Tserkovnyak, Yaroslav},
  journal = {Phys. Rev. Lett.},
  volume = {109},
  issue = {25},
  pages = {257002},
  numpages = {5},
  year = {2012},
  month = {Dec},
  publisher = {American Physical Society},
  doi = {10.1103/PhysRevLett.109.257002},
  url = {https://link.aps.org/doi/10.1103/PhysRevLett.109.257002}
}

@article{SchmidtNJP2013,
    doi = {10.1088/1367-2630/15/2/025043},
    url = {https://doi.org/10.1088/1367-2630/15/2/025043},
    year = {2013},
    month = {feb},
    publisher = {IOP Publishing},
    volume = {15},
    number = {2},
    pages = {025043},
    author = {Schmidt, Thomas L and Nunnenkamp, Andreas and Bruder, Christoph},
    title = {Microwave-controlled coupling of Majorana bound states},
    journal = {New J. Phys.}
}

@article{DmytrukPRB2015,
  title = {Cavity quantum electrodynamics with mesoscopic topological superconductors},
  author = {Dmytruk, Olesia and Trif, Mircea and Simon, Pascal},
  journal = {Phys. Rev. B},
  volume = {92},
  issue = {24},
  pages = {245432},
  numpages = {15},
  year = {2015},
  month = {Dec},
  publisher = {American Physical Society},
  doi = {10.1103/PhysRevB.92.245432},
  url = {https://link.aps.org/doi/10.1103/PhysRevB.92.245432}
}

@Article{ContaminNQI2021,
    author={Contamin, L. C.
    and Delbecq, M. R.
    and Dou{\c{c}}ot, B.
    and Cottet, A.
    and Kontos, T.},
    title={Hybrid light-matter networks of Majorana zero modes},
    journal={npj Quantum Inf},
    year={2021},
    month={Dec},
    day={16},
    volume={7},
    number={1},
    pages={171},
    issn={2056-6387},
    doi={10.1038/s41534-021-00508-w},
    url={https://doi.org/10.1038/s41534-021-00508-w}
}

@article{DmytrukPRB2023,
  title = {Microwave detection of gliding Majorana zero modes in nanowires},
  author = {Dmytruk, Olesia and Trif, Mircea},
  journal = {Phys. Rev. B},
  volume = {107},
  issue = {11},
  pages = {115418},
  numpages = {11},
  year = {2023},
  month = {Mar},
  publisher = {American Physical Society},
  doi = {10.1103/PhysRevB.107.115418},
  url = {https://link.aps.org/doi/10.1103/PhysRevB.107.115418}
}

@article{NguyenPRB2024,
  title = {Electron conductance and many-body marker of a cavity-embedded topological one-dimensional chain},
  author = {Nguyen, Danh-Phuong and Arwas, Geva and Ciuti, Cristiano},
  journal = {Phys. Rev. B},
  volume = {110},
  issue = {19},
  pages = {195416},
  numpages = {9},
  year = {2024},
  month = {Nov},
  publisher = {American Physical Society},
  doi = {10.1103/PhysRevB.110.195416},
  url = {https://link.aps.org/doi/10.1103/PhysRevB.110.195416}
}

@article{WangXPRB2019,
  title = {Cavity quantum electrodynamical Chern insulator: Towards light-induced quantized anomalous Hall effect in graphene},
  author = {Wang, Xiao and Ronca, Enrico and Sentef, Michael A.},
  journal = {Phys. Rev. B},
  volume = {99},
  issue = {23},
  pages = {235156},
  numpages = {7},
  year = {2019},
  month = {Jun},
  publisher = {American Physical Society},
  doi = {10.1103/PhysRevB.99.235156},
  url = {https://link.aps.org/doi/10.1103/PhysRevB.99.235156}
}

@article{GuerciPRL2020,
  title = {Superradiant Phase Transition in Electronic Systems and Emergent Topological Phases},
  author = {Guerci, Daniele and Simon, Pascal and Mora, Christophe},
  journal = {Phys. Rev. Lett.},
  volume = {125},
  issue = {25},
  pages = {257604},
  numpages = {6},
  year = {2020},
  month = {Dec},
  publisher = {American Physical Society},
  doi = {10.1103/PhysRevLett.125.257604},
  url = {https://link.aps.org/doi/10.1103/PhysRevLett.125.257604}
}

@article{loring.15,
    title = {K-theory and pseudospectra for topological insulators},
        journal = {Ann. Phys.},
    volume = {356},
    pages = {383-416},
    year = {2015},
    issn = {0003-4916},
    doi = {https://doi.org/10.1016/j.aop.2015.02.031},
    author = {Terry A. Loring}
}

@article{loring2017finitevolume,
      title={Finite volume calculation of $K$-theory invariants}, 
      author={Terry Loring and Hermann Schulz-Baldes},
      eprint={1701.07455},
      archivePrefix={arXiv}
}

@misc{loring.19,
    Author = {Terry A. Loring},
    Title = {A Guide to the Bott Index and Localizer Index},
    Year = {2019},
    Eprint = {arXiv:1907.11791},
}

@article{qi.na.24,
  title = {Real-space topological invariant for time-quasiperiodic Majorana modes},
  author = {Qi, Zihao and Na, Ilyoun and Refael, Gil and Peng, Yang},
  journal = {Phys. Rev. B},
  volume = {110},
  issue = {1},
  pages = {014309},
  numpages = {9},
  year = {2024},
  month = {Jul},
  publisher = {American Physical Society},
  doi = {10.1103/PhysRevB.110.014309},
  url = {https://link.aps.org/doi/10.1103/PhysRevB.110.014309}
}

@article{ghosh.arouca.24,
  title = {Local and energy-resolved topological invariants for Floquet systems},
  author = {Ghosh, Arnob Kumar and Arouca, Rodrigo and Black-Schaffer, Annica M.},
  journal = {Phys. Rev. B},
  volume = {110},
  issue = {24},
  pages = {245306},
  numpages = {11},
  year = {2024},
  month = {Dec},
  publisher = {American Physical Society},
  doi = {10.1103/PhysRevB.110.245306},
  url = {https://link.aps.org/doi/10.1103/PhysRevB.110.245306}
}

@misc{ghoshMAC2025,
      title={Laser-induced topological phases in monolayer amorphous carbon}, 
      author={Arnob Kumar Ghosh and Quentin Marsal and Annica M. Black-Schaffer},
      eprint={2508.09571}
}

@article{CerjanJMP2023,
    author = {Cerjan, Alexander and Loring, Terry A. and Vides, Fredy},
    title = "{Quadratic pseudospectrum for identifying localized states}",
    journal = {J. Math. Phys.},
    volume = {64},
    number = {2},
    pages = {023501},
    year = {2023},
    month = {02},
    issn = {0022-2488},
    doi = {10.1063/5.0098336},
    url = {https://doi.org/10.1063/5.0098336}
}

@article{WongPRB2023,
  title = {Probing topology in nonlinear topological materials using numerical $K$-theory},
  author = {Wong, Stephan and Loring, Terry A. and Cerjan, Alexander},
  journal = {Phys. Rev. B},
  volume = {108},
  issue = {19},
  pages = {195142},
  numpages = {12},
  year = {2023},
  month = {Nov},
  publisher = {American Physical Society},
  doi = {10.1103/PhysRevB.108.195142},
  url = {https://link.aps.org/doi/10.1103/PhysRevB.108.195142}
}

@article{thouless.74,
title = {Electrons in disordered systems and the theory of localization},
journal = {Phys. Rep.},
volume = {13},
number = {3},
pages = {93-142},
year = {1974},
issn = {0370-1573},
doi = {https://doi.org/10.1016/0370-1573(74)90029-5},
url = {https://www.sciencedirect.com/science/article/pii/0370157374900295},
author = {D.J. Thouless}
}

@article{Kitaev_2001,
	doi = {10.1070/1063-7869/44/10s/s29},
	url = {https://doi.org/10.1070/1063-7869/44/10s/s29},
	year = {2001},
	month = {oct},
	publisher = {Uspekhi Fizicheskikh Nauk ({UFN}) Journal},
	volume = {44},
	number = {10S},
	pages = {131--136},
	author = {A Yu Kitaev},
	title = {{Unpaired Majorana fermions in quantum wires}},
	journal = {Phys.-Usp.}
}

@Article{Li_16,
author={Li, Jian
and Neupert, Titus
and Bernevig, B. Andrei
and Yazdani, Ali},
title={Manipulating Majorana zero modes on atomic rings with an external magnetic field},
journal={Nature Communications},
year={2016},
month={Jan},
day={21},
volume={7},
number={1},
pages={10395},
issn={2041-1723},
doi={10.1038/ncomms10395},
url={https://doi.org/10.1038/ncomms10395}
}

@article{theiler_19,
  title = {Majorana bound state localization and energy oscillations for magnetic impurity chains on conventional superconductors},
  author = {Theiler, Andreas and Bj\"ornson, Kristofer and Black-Schaffer, Annica M.},
  journal = {Phys. Rev. B},
  volume = {100},
  issue = {21},
  pages = {214504},
  numpages = {7},
  year = {2019},
  month = {Dec},
  publisher = {American Physical Society},
  doi = {10.1103/PhysRevB.100.214504},
  url = {https://link.aps.org/doi/10.1103/PhysRevB.100.214504}
}

@article{qi2011topological,
  title = {{Topological insulators and superconductors}},
  author = {Qi, Xiao-Liang and Zhang, Shou-Cheng},
  journal = {Rev. Mod. Phys.},
  volume = {83},
  issue = {4},
  pages = {1057--1110},
  numpages = {0},
  year = {2011},
  month = {Oct},
  publisher = {American Physical Society},
  doi = {10.1103/RevModPhys.83.1057},
  url = {https://link.aps.org/doi/10.1103/RevModPhys.83.1057}
}

@article{Alicea_2012,
	doi = {10.1088/0034-4885/75/7/076501},
	url = {https://doi.org/10.1088/0034-4885/75/7/076501},
	year = 2012,
	month = {jun},
	publisher = {{IOP} Publishing},
	volume = {75},
	number = {7},
	pages = {076501},
	author = {Jason Alicea},
	title = {{New directions in the pursuit of Majorana fermions in solid state systems}},
	journal = {Rep. Prog. Phys.}
}

@article{Leijnse_2012,
	doi = {10.1088/0268-1242/27/12/124003},
	url = {https://doi.org/10.1088/0268-1242/27/12/124003},
	year = 2012,
	month = {nov},
	publisher = {{IOP} Publishing},
	volume = {27},
	number = {12},
	pages = {124003},
	author = {Martin Leijnse and Karsten Flensberg},
	title = {Introduction to topological superconductivity and Majorana fermions},
	journal = {Semicond. Sci. Technol.}
}

@article{LeijnsePM2012,
  title = {Parity qubits and poor man's Majorana bound states in double quantum dots},
  author = {Leijnse, Martin and Flensberg, Karsten},
  journal = {Phys. Rev. B},
  volume = {86},
  issue = {13},
  pages = {134528},
  numpages = {7},
  year = {2012},
  month = {Oct},
  publisher = {American Physical Society},
  doi = {10.1103/PhysRevB.86.134528},
  url = {https://link.aps.org/doi/10.1103/PhysRevB.86.134528}
}

@article{beenakker2013search,
  title={Search for Majorana fermions in superconductors},
  author={Beenakker, CWJ},
  journal={Annu. Rev. Condens. Matter Phys.},
  volume={4},
  number={1},
  pages={113--136},
  year={2013},
  doi = {10.1146/annurev-conmatphys-030212-184337},
  URL = {https://doi.org/10.1146/annurev-conmatphys-030212-184337},  
  publisher={Annual Reviews}
}

@article{ramonaquado2017,
  title={Majorana quasiparticles in condensed matter},
  author={Aguado, R.},
  journal={Riv. Nuovo Cimento},
  volume={40},
  number={11},
  pages={523},
  year={2017},
  doi={10.1393/ncr/i2017-10141-9},
  url={https://doi.org/10.1393/ncr/i2017-10141-9},
  publisher={IOP Publishing}
}

@article{tanaka2024theory,
    author = {Tanaka, Yukio and Tamura, Shun and Cayao, Jorge},
    title = {Theory of majorana zero modes in unconventional superconductors},
    journal = {Prog. Theor. Exp. Phys.},
    pages = {ptae065},
    year = {2024},
    month = {05},
    issn = {2050-3911},
    doi = {10.1093/ptep/ptae065},
    url = {https://doi.org/10.1093/ptep/ptae065}
}

@article{LutchynPRL2010,
  title = {{Majorana Fermions and a Topological Phase Transition in Semiconductor-Superconductor Heterostructures}},
  author = {Lutchyn, Roman M. and Sau, Jay D. and Das Sarma, S.},
  journal = {Phys. Rev. Lett.},
  volume = {105},
  issue = {7},
  pages = {077001},
  numpages = {4},
  year = {2010},
  month = {Aug},
  publisher = {American Physical Society},
  doi = {10.1103/PhysRevLett.105.077001},
  url = {https://link.aps.org/doi/10.1103/PhysRevLett.105.077001}
}

@article{Oreg2010,
  title = {Helical Liquids and Majorana Bound States in Quantum Wires},
  author = {Oreg, Yuval and Refael, Gil and von Oppen, Felix},
  journal = {Phys. Rev. Lett.},
  volume = {105},
  issue = {17},
  pages = {177002},
  numpages = {4},
  year = {2010},
  month = {Oct},
  publisher = {American Physical Society},
  doi = {10.1103/PhysRevLett.105.177002},
  url = {https://link.aps.org/doi/10.1103/PhysRevLett.105.177002}
}

@article{DengNano2012,
  author = {Deng, M. T. and Yu, C. L. and Huang, G. Y. and Larsson, M. and Caroff, P. and Xu, H. Q.},
  title = {Anomalous Zero-Bias Conductance Peak in a Nb–InSb Nanowire–Nb Hybrid Device},
  journal = {Nano Lett.},
  volume = {12},
  number = {12},
  pages = {6414-6419},
  year = {2012},
  doi = {10.1021/nl303758w},
  URL = {https://doi.org/10.1021/nl303758w}
}

@Article{Rokhinson2012,
	author={Rokhinson, Leonid P.
	and Liu, Xinyu
	and Furdyna, Jacek K.},
	title={The fractional a.c. Josephson effect in a semiconductor--superconductor nanowire as a signature of Majorana particles},
	journal={Nat. Phys.},
	year={2012},
	month={Nov},
	day={01},
	volume={8},
	number={11},
	pages={795-799},
	issn={1745-2481},
	doi={10.1038/nphys2429},
	url={https://doi.org/10.1038/nphys2429}
}

@Article{Albrecht2016,
	author={Albrecht, S. M.
	and Higginbotham, A. P.
	and Madsen, M.
	and Kuemmeth, F.
	and Jespersen, T. S.
	and Nyg{\aa}rd, J.
	and Krogstrup, P.
	and Marcus, C. M.},
	title={Exponential protection of zero modes in Majorana islands},
	journal={Nature},
	year={2016},
	month={Mar},
	day={01},
	volume={531},
	number={7593},
	pages={206-209},
	issn={1476-4687},
	doi={10.1038/nature17162},
	url={https://doi.org/10.1038/nature17162}
}

@article {Deng2016,
	author = {Deng, M. T. and Vaitiekenas, S. and Hansen, E. B. and Danon, J. and Leijnse, M. and Flensberg, K. and Nyg{\r a}rd, J. and Krogstrup, P. and Marcus, C. M.},
	title = {Majorana bound state in a coupled quantum-dot hybrid-nanowire system},
	volume = {354},
	number = {6319},
	pages = {1557--1562},
	year = {2016},
	doi = {10.1126/science.aaf3961},
	publisher = {American Association for the Advancement of Science},
	issn = {0036-8075},
	URL = {https://science.sciencemag.org/content/354/6319/1557},
	journal = {Science}
}

@article{JunSciAdv2017,
  author = {Jun Chen  and Peng Yu  and John Stenger  and Moïra Hocevar  and Diana Car  and Sébastien R. Plissard  and Erik P. A. M. Bakkers  and Tudor D. Stanescu  and Sergey M. Frolov },
  title = {Experimental phase diagram of zero-bias conductance peaks in superconductor/semiconductor nanowire devices},
  journal = {Sci. Adv.},
  volume = {3},
  number = {9},
  pages = {e1701476},
  year = {2017},
  doi = {10.1126/sciadv.1701476},
  URL = {https://www.science.org/doi/abs/10.1126/sciadv.1701476}
}

@Article{Gul2018,
  author={G{\"u}l, {\"O}nder
  and Zhang, Hao
  and Bommer, Jouri D. S.
  and de Moor, Michiel W. A.
  and Car, Diana
  and Plissard, S{\'e}bastien R.
  and Bakkers, Erik P. A. M.
  and Geresdi, Attila
  and Watanabe, Kenji
  and Taniguchi, Takashi
  and Kouwenhoven, Leo P.},
  title={Ballistic Majorana nanowire devices},
  journal={Nat. Nanotechnol.},
  year={2018},
  month={Mar},
  day={01},
  volume={13},
  number={3},
  pages={192-197},
  issn={1748-3395},
  doi={10.1038/s41565-017-0032-8},
  url={https://doi.org/10.1038/s41565-017-0032-8}
}

@article{das2012zero,
	doi = {https://doi.org/10.1038/nphys2479},
  url = {https://doi.org/10.1038/nphys2479},
  title={Zero-bias peaks and splitting in an {\rm Al}--{\rm InAs} nanowire topological superconductor as a signature of Majorana fermions},
  author={Das, Anindya and Ronen, Yuval and Most, Yonatan and Oreg, Yuval and Heiblum, Moty and Shtrikman, Hadas},
  journal={Nat. Phys.},
  volume={8},
  number={12},
  pages={887--895},
  year={2012},
  publisher={Nature Publishing Group}
}

@article{Mourik2012Science,
  author = {V. Mourik  and K. Zuo  and S. M. Frolov  and S. R. Plissard  and E. P. A. M. Bakkers  and L. P. Kouwenhoven },
  title = {Signatures of Majorana Fermions in Hybrid Superconductor-Semiconductor Nanowire Devices},
  journal = {Science},
  volume = {336},
  number = {6084},
  pages = {1003-1007},
  year = {2012},
  doi = {10.1126/science.1222360},
  URL = {https://www.science.org/doi/abs/10.1126/science.1222360}
}

@article{NichelePRL2017,
  title = {Scaling of Majorana Zero-Bias Conductance Peaks},
  author = {Nichele, Fabrizio and Drachmann, Asbj\o{}rn C. C. and Whiticar, Alexander M. and O'Farrell, Eoin C. T. and Suominen, Henri J. and Fornieri, Antonio and Wang, Tian and Gardner, Geoffrey C. and Thomas, Candice and Hatke, Anthony T. and Krogstrup, Peter and Manfra, Michael J. and Flensberg, Karsten and Marcus, Charles M.},
  journal = {Phys. Rev. Lett.},
  volume = {119},
  issue = {13},
  pages = {136803},
  numpages = {5},
  year = {2017},
  month = {Sep},
  publisher = {American Physical Society},
  doi = {10.1103/PhysRevLett.119.136803},
  url = {https://link.aps.org/doi/10.1103/PhysRevLett.119.136803}
}

@Article{Zhang2017NatCommun,
  author={Zhang, Hao
  and G{\"u}l, {\"O}nder
  and Conesa-Boj, Sonia
  and Nowak, Micha{\l} P.
  and Wimmer, Michael
  and Zuo, Kun
  and Mourik, Vincent
  and de Vries, Folkert K.
  and van Veen, Jasper
  and de Moor, Michiel W. A.
  and Bommer, Jouri D. S.
  and van Woerkom, David J.
  and Car, Diana
  and Plissard, S{\'e}bastien R.
  and Bakkers, Erik P.A.M.
  and Quintero-P{\'e}rez, Marina
  and Cassidy, Maja C.
  and Koelling, Sebastian
  and Goswami, Srijit
  and Watanabe, Kenji
  and Taniguchi, Takashi
  and Kouwenhoven, Leo P.},
  title={Ballistic superconductivity in semiconductor nanowires},
  journal={Nat. Commun.},
  year={2017},
  month={Jul},
  day={06},
  volume={8},
  number={1},
  pages={16025},
  issn={2041-1723},
  doi={10.1038/ncomms16025},
  url={https://doi.org/10.1038/ncomms16025}
}

@Article{Grivnin2019,
  author={Grivnin, Anna
  and Bor, Ella
  and Heiblum, Moty
  and Oreg, Yuval
  and Shtrikman, Hadas},
  title={Concomitant opening of a bulk-gap with an emerging possible Majorana zero mode},
  journal={Nat. Commun.},
  year={2019},
  month={Apr},
  day={29},
  volume={10},
  number={1},
  pages={1940},
  issn={2041-1723},
  doi={10.1038/s41467-019-09771-0},
  url={https://doi.org/10.1038/s41467-019-09771-0}
}

@article{ChenPRL2019,
  title = {Ubiquitous Non-Majorana Zero-Bias Conductance Peaks in Nanowire Devices},
  author = {Chen, J. and Woods, B. D. and Yu, P. and Hocevar, M. and Car, D. and Plissard, S. R. and Bakkers, E. P. A. M. and Stanescu, T. D. and Frolov, S. M.},
  journal = {Phys. Rev. Lett.},
  volume = {123},
  issue = {10},
  pages = {107703},
  numpages = {6},
  year = {2019},
  month = {Sep},
  publisher = {American Physical Society},
  doi = {10.1103/PhysRevLett.123.107703},
  url = {https://link.aps.org/doi/10.1103/PhysRevLett.123.107703}
}

@article{Finck2013,
  title = {Anomalous Modulation of a Zero-Bias Peak in a Hybrid Nanowire-Superconductor Device},
  author = {Finck, A. D. K. and Van Harlingen, D. J. and Mohseni, P. K. and Jung, K. and Li, X.},
  journal = {Phys. Rev. Lett.},
  volume = {110},
  issue = {12},
  pages = {126406},
  numpages = {5},
  year = {2013},
  month = {Mar},
  publisher = {American Physical Society},
  doi = {10.1103/PhysRevLett.110.126406},
  url = {https://link.aps.org/doi/10.1103/PhysRevLett.110.126406}
}

@Article{Dvir23,
author={Dvir, Tom and Wang, Guanzhong and van Loo, Nick and Liu, Chun-Xiao and Mazur, Grzegorz P. and Bordin, Alberto and ten Haaf, Sebastiaan L. D. and Wang, Ji-Yin and van Driel, David and Zatelli, Francesco and Li, Xiang and Malinowski, Filip K. and Gazibegovic, Sasa and Badawy, Ghada and Bakkers, Erik P. A. M. and Wimmer, Michael and Kouwenhoven, Leo P.},
title={Realization of a minimal Kitaev chain in coupled quantum dots},
journal={Nature},
year={2023},
month={Feb},
day={01},
volume={614},
number={7948},
pages={445-450},
issn={1476-4687},
doi={10.1038/s41586-022-05585-1},
url={https://doi.org/10.1038/s41586-022-05585-1}
}

@Article{Haaf24,
author={ten Haaf, Sebastiaan L. D. and Wang, Qingzhen and Bozkurt, A. Mert and Liu, Chun-Xiao and Kulesh, Ivan and Kim, Philip and Xiao, Di and Thomas, Candice and Manfra, Michael J. and Dvir, Tom and Wimmer, Michael and Goswami, Srijit},
title={A two-site Kitaev chain in a two-dimensional electron gas},
journal={Nature},
year={2024},
month={Jun},
day={01},
volume={630},
number={8016},
pages={329-334},
issn={1476-4687},
doi={10.1038/s41586-024-07434-9},
url={https://doi.org/10.1038/s41586-024-07434-9}
}

@Article{Zatelli24,
author={Zatelli, Francesco and van Driel, David and Xu, Di and Wang, Guanzhong and Liu, Chun-Xiao and Bordin, Alberto and Roovers, Bart and Mazur, Grzegorz P. and van Loo, Nick and Wolff, Jan C. and Bozkurt, A. Mert and Badawy, Ghada and Gazibegovic, Sasa and Bakkers, Erik P. A. M. and Wimmer, Michael and Kouwenhoven, Leo P. and Dvir, Tom},
title={Robust poor man's Majorana zero modes using Yu-Shiba-Rusinov states},
journal={Nat. Commun.},
year={2024},
month={Sep},
day={11},
volume={15},
number={1},
pages={7933},
issn={2041-1723},
doi={10.1038/s41467-024-52066-2},
url={https://doi.org/10.1038/s41467-024-52066-2}
}

@article{Ivanov2001,
  title = {{Non-Abelian Statistics of Half-Quantum Vortices in $\mathit{p}$-Wave Superconductors}},
  author = {Ivanov, D. A.},
  journal = {Phys. Rev. Lett.},
  volume = {86},
  issue = {2},
  pages = {268--271},
  numpages = {0},
  year = {2001},
  month = {Jan},
  publisher = {American Physical Society},
  doi = {10.1103/PhysRevLett.86.268},
  url = {https://link.aps.org/doi/10.1103/PhysRevLett.86.268}
}

@article{freedman2003topological,
  title={Topological quantum computation},
  author={Freedman, Michael and Kitaev, Alexei and Larsen, Michael and Wang, Zhenghan},
  journal={Bull. Amer. Math. Soc.},
  volume={40},
  number={1},
  pages={31--38},
  year={2003},
  doi={10.1090/S0273-0979-02-00964-3}
}

@article{NayakRMP2008,
  title = {Non-Abelian anyons and topological quantum computation},
  author = {Nayak, Chetan and Simon, Steven H. and Stern, Ady and Freedman, Michael and Das Sarma, Sankar},
  journal = {Rev. Mod. Phys.},
  volume = {80},
  issue = {3},
  pages = {1083--1159},
  numpages = {0},
  year = {2008},
  month = {Sep},
  publisher = {American Physical Society},
  doi = {10.1103/RevModPhys.80.1083},
  url = {https://link.aps.org/doi/10.1103/RevModPhys.80.1083}
}

@article{JiangPRB2025,
  title = {Angular momentum dependent spectral shift in chiral vacuum cavities},
  author = {Jiang, Qing-Dong},
  journal = {Phys. Rev. B},
  volume = {111},
  issue = {20},
  pages = {205405},
  numpages = {8},
  year = {2025},
  month = {May},
  publisher = {American Physical Society},
  doi = {10.1103/PhysRevB.111.205405},
  url = {https://link.aps.org/doi/10.1103/PhysRevB.111.205405}
}

@article{yangCommPhys2025,
	title = {Emergent {Haldane} model and photon-valley locking in chiral cavities},
	author = {Yang, Liu and Jiang, Qing-Dong},
	journal = {Commun. Phys.},
	volume = {8},
	issn = {2399-3650},
	doi = {10.1038/s42005-025-02060-x},
	url = {https://www.nature.com/articles/s42005-025-02060-x},
	number = {1},
	urldate = {2025-11-06},
	year = {2025},
	note = {Publisher: Nature Publishing Group},
	pages = {126}
}

@misc{CardosoArXiV2025,
	title = {Cavity {Quantum} {Hall} {Hydrodynamics}},
	archivePrefix = {arXiv},
	author = {Cardoso, Gabriel and Yang, Liu and Hansson, Thors Hans and Jiang, Qing-Dong},
	year = {2025},
	eprint = {arXiv:2501.01492},
}

@misc{YangArxiV2025,
	title = {Quantum {Hall} effect in a chiral cavity},
	archivePrefix = {arXiv},
	author = {Yang, Liu and Cardoso, Gabriel and Hansson, Thors Hans and Jiang, Qing-Dong},
	year = {2025},
	eprint = {2503.11757}
}

@article{weiArxiV2025,
	title = {Cavity-{Vacuum}-{Induced} {Chiral} {Spin} {Liquids} in {Kagome} {Lattices}: {Tuning} and {Probing} {Topological} {Quantum} {Phases} via {Cavity} {Quantum} {Electrodynamics}},
	archivePrefix = {arXiv},
	author = {Wei, Chenan and Yang, Liu and Jiang, Qing-Dong},
	eprint = {2411.08121},
    year = {2024}
}

@article{akhmerov09,
  title = {Electrically Detected Interferometry of Majorana Fermions in a Topological Insulator},
  author = {Akhmerov, A. R. and Nilsson, Johan and Beenakker, C. W. J.},
  journal = {Phys. Rev. Lett.},
  volume = {102},
  issue = {21},
  pages = {216404},
  numpages = {4},
  year = {2009},
  month = {May},
  publisher = {American Physical Society},
  doi = {10.1103/PhysRevLett.102.216404},
  url = {https://link.aps.org/doi/10.1103/PhysRevLett.102.216404}
}

@article{grosfeld11,
author = {Eytan Grosfeld  and Ady Stern },
title = {Observing Majorana bound states of Josephson vortices in topological superconductors},
journal = {PNAS},
volume = {108},
number = {29},
pages = {11810-11814},
year = {2011},
doi = {10.1073/pnas.1101469108},
URL = {https://www.pnas.org/doi/abs/10.1073/pnas.1101469108}
}

@article{vaezi13,
  title = {Fractional topological superconductor with fractionalized Majorana fermions},
  author = {Vaezi, Abolhassan},
  journal = {Phys. Rev. B},
  volume = {87},
  issue = {3},
  pages = {035132},
  numpages = {13},
  year = {2013},
  month = {Jan},
  publisher = {American Physical Society},
  doi = {10.1103/PhysRevB.87.035132},
  url = {https://link.aps.org/doi/10.1103/PhysRevB.87.035132}
}

@article{sticlet14,
  title = {From fractionally charged solitons to Majorana bound states in a one-dimensional interacting model},
  author = {Sticlet, Doru and Seabra, Luis and Pollmann, Frank and Cayssol, J\'er\^ome},
  journal = {Phys. Rev. B},
  volume = {89},
  issue = {11},
  pages = {115430},
  numpages = {18},
  year = {2014},
  month = {Mar},
  publisher = {American Physical Society},
  doi = {10.1103/PhysRevB.89.115430},
  url = {https://link.aps.org/doi/10.1103/PhysRevB.89.115430}
}

@article{anappara09,
  title = {Signatures of the ultrastrong light-matter coupling regime},
  author = {Anappara, Aji A. and De Liberato, Simone and Tredicucci, Alessandro and Ciuti, Cristiano and Biasiol, Giorgio and Sorba, Lucia and Beltram, Fabio},
  journal = {Phys. Rev. B},
  volume = {79},
  issue = {20},
  pages = {201303},
  numpages = {4},
  year = {2009},
  month = {May},
  publisher = {American Physical Society},
  doi = {10.1103/PhysRevB.79.201303},
  url = {https://link.aps.org/doi/10.1103/PhysRevB.79.201303}
}

@article{scalari12,
author = {G. Scalari  and C. Maissen  and D. Turčinková  and D. Hagenmüller  and S. De Liberato  and C. Ciuti  and C. Reichl  and D. Schuh  and W. Wegscheider  and M. Beck  and J. Faist },
title = {Ultrastrong Coupling of the Cyclotron Transition of a 2D Electron Gas to a THz Metamaterial},
journal = {Science},
volume = {335},
number = {6074},
pages = {1323-1326},
year = {2012},
doi = {10.1126/science.1216022},
URL = {https://www.science.org/doi/abs/10.1126/science.1216022}
}

@article{maissen14,
  title = {Ultrastrong coupling in the near field of complementary split-ring resonators},
  author = {Maissen, Curdin and Scalari, Giacomo and Valmorra, Federico and Beck, Mattias and Faist, J\'er\^ome and Cibella, Sara and Leoni, Roberto and Reichl, Christian and Charpentier, Christophe and Wegscheider, Werner},
  journal = {Phys. Rev. B},
  volume = {90},
  issue = {20},
  pages = {205309},
  numpages = {9},
  year = {2014},
  month = {Nov},
  publisher = {American Physical Society},
  doi = {10.1103/PhysRevB.90.205309},
  url = {https://link.aps.org/doi/10.1103/PhysRevB.90.205309}
}

@Article{Gambino2014,
author={Gambino, Salvatore
and Mazzeo, Marco
and Genco, Armando
and Di Stefano, Omar
and Savasta, Salvatore
and Patan{\`e}, Salvatore
and Ballarini, Dario
and Mangione, Federica
and Lerario, Giovanni
and Sanvitto, Daniele
and Gigli, Giuseppe},
title={Exploring Light--Matter Interaction Phenomena under Ultrastrong Coupling Regime},
journal={ACS Photonics},
year={2014},
month={Oct},
day={15},
publisher={American Chemical Society},
volume={1},
number={10},
pages={1042-1048},
doi={10.1021/ph500266d},
url={https://doi.org/10.1021/ph500266d}
}

@Article{Chikkaraddy2016,
author={Chikkaraddy, Rohit
and de Nijs, Bart
and Benz, Felix
and Barrow, Steven J.
and Scherman, Oren A.
and Rosta, Edina
and Demetriadou, Angela
and Fox, Peter
and Hess, Ortwin
and Baumberg, Jeremy J.},
title={Single-molecule strong coupling at room temperature in plasmonic nanocavities},
journal={Nature},
year={2016},
month={Jul},
day={01},
volume={535},
number={7610},
pages={127-130},
issn={1476-4687},
doi={10.1038/nature17974},
url={https://doi.org/10.1038/nature17974}
}

@Article{Yoshihara2017,
author={Yoshihara, Fumiki
and Fuse, Tomoko
and Ashhab, Sahel
and Kakuyanagi, Kosuke
and Saito, Shiro
and Semba, Kouichi},
title={Superconducting qubit--oscillator circuit beyond the ultrastrong-coupling regime},
journal={Nature Physics},
year={2017},
month={Jan},
day={01},
volume={13},
number={1},
pages={44-47},
issn={1745-2481},
doi={10.1038/nphys3906},
url={https://doi.org/10.1038/nphys3906}
}

@article{flick17,
author = {Johannes Flick  and Michael Ruggenthaler  and Heiko Appel  and Angel Rubio },
title = {Atoms and molecules in cavities, from weak to strong coupling in quantum-electrodynamics (QED) chemistry},
journal = {PNAS},
volume = {114},
number = {12},
pages = {3026-3034},
year = {2017},
doi = {10.1073/pnas.1615509114},
URL = {https://www.pnas.org/doi/abs/10.1073/pnas.1615509114}
}

@Article{Bayer2017,
author={Bayer, Andreas
and Pozimski, Marcel
and Schambeck, Simon
and Schuh, Dieter
and Huber, Rupert
and Bougeard, Dominique
and Lange, Christoph},
title={Terahertz Light--Matter Interaction beyond Unity Coupling Strength},
journal={Nano Letters},
year={2017},
month={Oct},
day={11},
publisher={American Chemical Society},
volume={17},
number={10},
pages={6340-6344},
issn={1530-6984},
doi={10.1021/acs.nanolett.7b03103},
url={https://doi.org/10.1021/acs.nanolett.7b03103}
}

@article{genco17,
author = {Genco, Armando and Ridolfo, Alessandro and Savasta, Salvatore and Patanè, Salvatore and Gigli, Giuseppe and Mazzeo, Marco},
title = {Bright Polariton Coumarin-Based OLEDs Operating in the Ultrastrong Coupling Regime},
journal = {Adv. Opt. Mater.},
volume = {6},
number = {17},
pages = {1800364},
doi = {https://doi.org/10.1002/adom.201800364},
url = {https://advanced.onlinelibrary.wiley.com/doi/abs/10.1002/adom.201800364},
year = {2018}
}

@Article{FriskKockum2019,
author={Frisk Kockum, Anton
and Miranowicz, Adam
and De Liberato, Simone
and Savasta, Salvatore
and Nori, Franco},
title={Ultrastrong coupling between light and matter},
journal={Nature Reviews Physics},
year={2019},
month={Jan},
day={01},
volume={1},
number={1},
pages={19-40},
issn={2522-5820},
doi={10.1038/s42254-018-0006-2},
url={https://doi.org/10.1038/s42254-018-0006-2}
}

@Article{Halbhuber2020,
author={Halbhuber, M.
and Mornhinweg, J.
and Zeller, V.
and Ciuti, C.
and Bougeard, D.
and Huber, R.
and Lange, C.},
title={Non-adiabatic stripping of a cavity field from electrons in the deep-strong coupling regime},
journal={Nature Photonics},
year={2020},
month={Nov},
day={01},
volume={14},
number={11},
pages={675-679},
issn={1749-4893},
doi={10.1038/s41566-020-0673-2},
url={https://doi.org/10.1038/s41566-020-0673-2}
}

@article{akhmerov11,
  title = {Quantized Conductance at the Majorana Phase Transition in a Disordered Superconducting Wire},
  author = {Akhmerov, A. R. and Dahlhaus, J. P. and Hassler, F. and Wimmer, M. and Beenakker, C. W. J.},
  journal = {Phys. Rev. Lett.},
  volume = {106},
  issue = {5},
  pages = {057001},
  numpages = {4},
  year = {2011},
  month = {Jan},
  publisher = {American Physical Society},
  doi = {10.1103/PhysRevLett.106.057001},
  url = {https://link.aps.org/doi/10.1103/PhysRevLett.106.057001}
}

@article{sau13,
  title = {Density of states of disordered topological superconductor-semiconductor hybrid nanowires},
  author = {Sau, Jay D. and Das Sarma, S.},
  journal = {Phys. Rev. B},
  volume = {88},
  issue = {6},
  pages = {064506},
  numpages = {7},
  year = {2013},
  month = {Aug},
  publisher = {American Physical Society},
  doi = {10.1103/PhysRevB.88.064506},
  url = {https://link.aps.org/doi/10.1103/PhysRevB.88.064506}
}

@article{awoga17,
  title = {Disorder robustness and protection of Majorana bound states in ferromagnetic chains on conventional superconductors},
  author = {Awoga, Oladunjoye A. and Bj\"ornson, Kristofer and Black-Schaffer, Annica M.},
  journal = {Phys. Rev. B},
  volume = {95},
  issue = {18},
  pages = {184511},
  numpages = {6},
  year = {2017},
  month = {May},
  publisher = {American Physical Society},
  doi = {10.1103/PhysRevB.95.184511},
  url = {https://link.aps.org/doi/10.1103/PhysRevB.95.184511}
}

@article{kobialka20,
  title = {Dimerization-induced topological superconductivity in a Rashba nanowire},
  author = {Kobia\l{}ka, Aksel and Sedlmayr, Nicholas and Ma\ifmmode \acute{s}\else \'{s}\fi{}ka, Maciej M. and Doma\ifmmode \acute{n}\else \'{n}\fi{}ski, Tadeusz},
  journal = {Phys. Rev. B},
  volume = {101},
  issue = {8},
  pages = {085402},
  numpages = {8},
  year = {2020},
  month = {Feb},
  publisher = {American Physical Society},
  doi = {10.1103/PhysRevB.101.085402},
  url = {https://link.aps.org/doi/10.1103/PhysRevB.101.085402}
}

\end{document}